\documentclass[aps,prl,superscriptaddress,floatfix,showpacs,notitlepage,twocolumn]{revtex4-1}
\usepackage[latin9]{inputenc}
\usepackage{color}
\usepackage{textcomp}
\usepackage{amsmath}
\usepackage{amssymb}
\usepackage{mathtools}
\usepackage{graphicx}
\usepackage{esint}
\usepackage{framed} 
\usepackage{xcolor}
\usepackage{hyperref}
\usepackage{enumerate}
\usepackage{enumitem}
\usepackage{moreenum}
\hypersetup{
  linkcolor = black,
  citecolor  = blue,
  urlcolor = black,
  colorlinks = true,
}
\makeatletter

\usepackage{amsfonts}
\usepackage{bbold}
\usepackage{natbib}
\usepackage{svg}
\usepackage{float}

\hypersetup{    colorlinks,    linkcolor={blue!50!black},    citecolor={blue!50!black},    urlcolor={blue!80!black}}
\usepackage[all]{hypcap}
\usepackage{bbold}
\graphicspath{{img/}}

\newcommand*{\bra}[1]{\ensuremath{\langle #1 \vert}}
\newcommand*{\ket}[1]{\ensuremath{\vert #1 \rangle}}
\newcommand{\braket}[2]{\langle #1 | #2 \rangle}

\newcommand*{\tr}[1]{\mathrm{tr}\left(#1\right)}
\newcommand*{\ptr}[2]{\mathrm{tr}_{#1}\left(#2\right)}

\newcommand{\mc}[1]{\mathcal{#1}}

\newcommand{\lr}[1]{\left( #1 \right)}

\newcommand{\smean}[1]{\langle\langle#1\rangle\rangle}
\renewcommand{\vec}[1]{\textbf{#1}}
\renewcommand{\v}[1]{\textbf{#1}}
\newcommand{\vac}{\ket{\rm vac}}

\begin{document}

\title{{Entanglement-Optimal Trajectories of Many-Body Quantum  Markov Processes}}
\author{Tatiana Vovk}
\affiliation{Institute for Theoretical Physics, University of Innsbruck, 6020 Innsbruck, Austria}
\affiliation{Institute for Quantum Optics and Quantum Information of the Austrian Academy of Sciences, 6020 Innsbruck, Austria}
\author{Hannes Pichler}
\email{hannes.pichler@uibk.ac.at}
\affiliation{Institute for Theoretical Physics, University of Innsbruck, 6020 Innsbruck, Austria}
\affiliation{Institute for Quantum Optics and Quantum Information of the Austrian Academy of Sciences, 6020 Innsbruck, Austria}

\begin{abstract}
We develop a novel approach aimed at solving the equations of motion of open quantum many-body systems. It is based on a combination of generalized wave function trajectories and matrix product states. We introduce an adaptive quantum stochastic propagator, which minimizes the expected entanglement in the many-body quantum state, thus minimizing the computational cost of the matrix product state representation of each trajectory. We illustrate this approach on the example of a one-dimensional open Brownian circuit. We show that this model displays an entanglement phase transition between area and volume law when changing between different propagators and that our method autonomously finds an efficiently representable area law unravelling.
\end{abstract}
\maketitle

Classical simulation of the evolution of quantum many-body systems is a formidably hard task, in particular if the system is fully coherent \cite{preskillQuantumComputingNISQ2018a}. Most near-term intermediate scale quantum devices are however noisy, which opens a possibility for the existence of efficient classical algorithms for simulating the corresponding open system dynamics. Nonetheless it is often unclear how to best exploit this potential.

\begin{figure}[htb]\centering
\includegraphics[width=\linewidth]{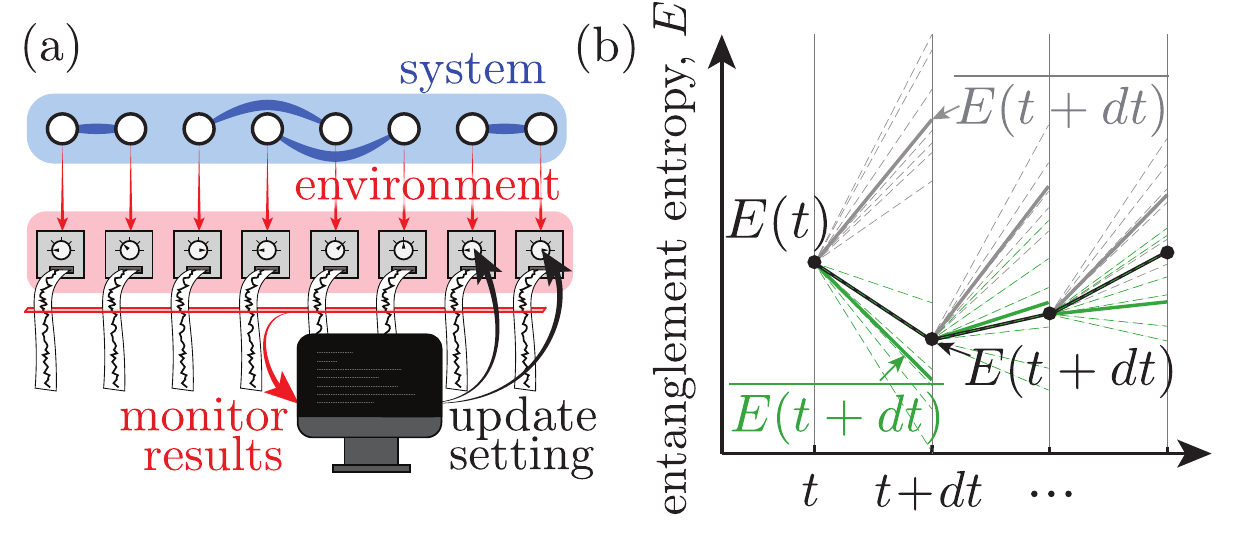}
\caption{(a) Schematic setting of a continuously monitored quantum many-body system with adaptive measurements. (b) Propagation of a state from time $t$ into the future with two options for the stochastic propagator (color coded, grey \textit{vs.}~green). The colored dashed lines are potential QTs for a given unravelling, their stochastic averages are shown as solid lines. A QT (black) is generated randomly at each time step, corresponding to the stochastic propagator that gives the lowest expected entanglement (green).
}
\label{fig:Fig1}
\end{figure}

Here we address this challenge by developing an algorithm that explicitly harnesses the quantum noise inherent to an open quantum system to minimize the computational cost of representing the many-body state. Our approach is based on a combination of matrix product states (MPSs) \cite{whiteDensityMatrixFormulation1992,vidalEfficientClassicalSimulation2003,schollwock2011density} and a generalization of the quantum trajectory (QT) method~\cite{dum1992monte, dalibard1992wave,gardiner1992wave,tian1992quantum,castin1993wave,van1998quantum}. The latter identifies the dynamics of an open quantum system as a stochastic evolution of pure quantum states,  corresponding to a continuous measurement of the environment~\cite{gardiner2015quantum}. Importantly, the choice of the monitored environment observables results in qualitatively  different ensembles of QTs. Our method utilizes this flexibility and continuously optimizes the monitored environment observables by predicting and minimizing the expected entanglement in the trajectory wave function (Fig.~\ref{fig:Fig1}), thus minimizing the computational cost of MPS representations.

We illustrate our approach by applying it to solve the Markovian master equation (ME) of an open Brownian circuit, where the coherent part of the evolution rapidly generates entanglement, while the dissipative part leads to dephasing. We show that various types of QT methods lead to ensembles that differ dramatically in their entanglement properties. This includes a phase transition between area law and volume law entangled ensembles, depending on the monitored environment observables. In addition to being an interesting phenomenon \textit{per~se}, with connections to recently discussed measurement-induced phase transitions \cite{liQuantumZenoEffect2018,liMeasurementdrivenEntanglementTransition2019,skinnerMeasurementInducedPhaseTransitions2019,chanUnitaryprojectiveEntanglementDynamics2019, gullansScalableProbesMeasurementinduced2020,choiQuantumErrorCorrection2020,bao2020theory,zabaloCriticalPropertiesMeasurementinduced2020,ippolitiEntanglementPhaseTransitions2021,zabaloOperatorScalingDimensions2021,botzungEngineeredDissipationInduced2021,mullerMeasurementinducedDarkState2021}, it provides an ideal test bed for our algorithm: We show that our entanglement predictor allows to generate ensembles of QTs that keep the system in the area law phase at all times. 

We are interested in open quantum many-body systems with $n$ constituents and $m$ dissipative channels described by a ME of a generic Lindblad form ($\hbar~=~1$)~\cite{lindblad1976generators}:
\begin{equation}
    \frac{d\rho}{dt} = -i \left[H_\mathrm{sys}, \rho\right] + \sum_{j=1}^m \gamma_j \left( c_j \rho c_j^\dagger - \frac{1}{2} \left\{\rho, c_j^\dagger c_j\right\}\right). 
    \label{eq:ME}
\end{equation}
Here, $\rho$ is the (many-body) density operator of the quantum system, $H_{\rm sys}$ is the (so far unspecified, potentially time-dependent) system Hamiltonian, and the $c_j$'s are the jump operators corresponding to the $j$-th decay channel with associated decay rate $\gamma_j$. In the following it is convenient to assume that $H_{\rm sys}$ is short-range interacting and the jump operators are local. Such a ME describes the reduced dynamics of a system coupled to an environment (or bath). The underlying joint evolution of system and environment is unitary and is described by the Schr\"odinger equation $i\partial_t\ket{\Psi(t)}=H(t)\ket{\Psi(t)}$, where $\ket{\Psi(t)}$ represents the joint state of the system and environment, and the Hamiltonian is given by~\cite{gardiner2015quantum}
\begin{align}\label{eq:QSSE}
H(t)=H_{\rm sys}+i \sum_{j=1}^m \sqrt{\gamma_j}\left[b_j(t)^\dag c_j-c_j^\dag b_j(t)\right].
\end{align}
Here, $b_j(t)$'s are so-called quantum noise operators with bosonic statistics $[b_j(t),b_{j'}(t')^\dag]=\delta_{j,j'}\delta(t-t')$. Those operators act on the degrees of freedom of the environment, which are initially in the vacuum state: $b_j(t)\ket{\Psi(0)} =0$. The ME \eqref{eq:ME} is obtained from the full Schr\"odinger equation by tracing out these bosonic degrees of freedom~\cite{gardiner2015quantum}.

The QT method solves the ME \eqref{eq:ME} by simulating a continuous monitoring of the environment degrees of freedom in \eqref{eq:QSSE}. It stochastically generates an ensemble of pure states of the system, called \textit{quantum trajectories}, from which the density matrix dynamics can be retrieved as a statistical average:
\begin{align}
    \rho(t) = \lim_{N\rightarrow \infty} \frac{1}{N}\sum_{k=1}^N \ket{\phi^{(k)}(t)} \bra{\phi^{(k)}(t)}.
    \label{eq:DM}
\end{align}
Here the quantum trajectories, $\ket{\phi^{(k)}(t)}$, represent the conditional state of the system for a particular (simulated) measurement history of the monitored environment (enumerated by $k$). They satisfy a stochastic equation of motion, i.e.~a quantum stochastic Schr\"odinger equation (QSSE)~\cite{gardiner2015quantum}. The form of the QSSE depends on the environment monitoring schemes, in particular on the monitored observables. For instance, the standard QT method, known as quantum jump approach, is obtained by continuous monitoring of the photon numbers, $b_j^\dag(t)b_j(t)$~\cite{dum1992monte, dalibard1992wave,gardiner1992wave, castin1993wave}.
Another approach is based on the balanced homodyne measurement, where a phase-sensitive homodyne current $b_j(t) e^{i\varphi_j} + b_j^\dagger(t) e^{-i\varphi_j}$ is continuously measured~\cite{tian1992quantum}. Here, the $\varphi_j$ specifies the measured quadrature.

A QT is generated by integrating the QSSE in (small) time steps, $dt$. In practice, one Trotterizes the corresponding propagator into a stochastic and a deterministic part. The stochastic component can be further split into independent components corresponding to $m$ decay channels~\cite{daley2014quantum}.  Specifically, a single integration step $\ket{\phi^{(k)}(t)}\rightarrow \ket{\phi^{(k)}(t+dt)}$ is achieved by a sequential application of stochastic operators $K_j$ for each decay channel $j$: Starting from the state $\ket{\phi^{(k)}_1(t)}=\ket{\phi^{(k)}(t)}$, one generates the sequence of normalized states
\begin{align}
    \ket{\phi^{(k)}_{j+1}(t)}&\propto K_j \ket{\phi^{(k)}_{j}(t)}, \quad (j=1,\dots ,m)
    \label{eq:propag}
\end{align}
and then concludes the integration step with the unitary operation $\ket{\phi^{(k)}(t+dt)}= e^{-i H_\mathrm{sys} dt} \ket{\phi^{(k)}_{m+1}(t)}.$
The form of the stochastic operator $K_j$ depends on the type of (simulated) environment measurement. For example, the operators simulating a homodyne measurement of the output field in channel $j$ is 
\begin{align}
    K_j^\mathrm{hom}=&e^{-\gamma_j dt c_j^\dag c_j/2} +\sqrt{\gamma_j} c_j e^{i\varphi_j}d\xi_j(t),\label{eq:propag_hom}
\end{align}
where $d\xi_j(t)= \sqrt{\gamma_j}\bra{\phi^{(k)}_{j}(t)} c_j e^{i\varphi_j}\! +  c_j^\dag e^{-i\varphi_j} \ket{\phi^{(k)}_{j}(t)}dt+dW_j(t)$ and the $d W_j (t)$'s are independent, normally distributed Gaussian variable with zero mean and variance $dt$, also known as Wiener increment. For a simulated number measurement, the  propagator is probabilistically selected from the two options:
 \begin{align}
    K_j^\mathrm{num} = 
    \left[
    \begin{array}{ll}
        \sqrt{\gamma_j dt}c_j &\mathrm{~with~ probability~} p_j, \\
        e^{-\gamma_j dt c_j^\dag c_j/2} &\mathrm{~with~probability~} 1- p_j,
    \end{array}
    \right .
\end{align}
where $p_j = \gamma_j dt\bra{\phi^{(k)}_{j}(t)}c_j^\dag c_j\ket{\phi^{(k)}_{j}(t)}$ corresponds to the probability of measuring a photon in channel $j$ between $t$ and $t+dt$.

While these two schemes are by far the most common ones, we stress that a multitude of stochastic propagators can be constructed by choosing different measurements of the bath operators specified via the eigenbases of arbitrary Hermitian functions of the bath operators, $X_j(t) = f[b_j(t), b_j^\dag (t)]$. This also includes measurement strategies that change with time and depend on prior measurement outcomes~\footnote{Since the trajectories are by construction sampled according to Born's rule, all such strategies, including the adaptive time- and history-dependent ones, generate trajectories that provide valid solutions of the ME [see Section~I of the Supplementary Materials (SM) for more detail].}. Importantly, the stochastic averages over linear functionals of the state projectors $\phi^{(k)}=\ket{\phi^{(k)}}\bra{\phi^{(k)}}$ do not depend on how the environment is monitored. This includes in particular the density operator \eqref{eq:DM} and the expectation values of linear observables. However the ensemble of QTs itself, as well as non-linear functionals of the trajectories, \textit{do depend} on the unravelling.
In particular this holds for the entanglement entropy (EE), which is of central interest in our work. We consider the EE between two partitions of the many-body system, $A \cup B$. In this case the EE of a single QT $\ket{\phi^{(k)}}$ is defined as the von Neumann entropy of the reduced state of one of the subsystems \{$\rho^{(k)}_A=\mathrm{tr}_B\left[{\phi^{(k)}}\right]$\}:
\begin{align}
    E\left[\phi^{(k)}\right]=S\left[\rho^{(k)}_A\right]=-\mathrm{tr}\left[{\rho^{(k)}_A\log_2 \rho^{(k)}_A}\right].
\end{align}
The ensemble averaged EE (EAEE) is then defined as 
\begin{align}
    \overline E = \lim_{N\rightarrow \infty}\frac{1}{N}\sum_{k=1}^N E\left[\phi^{(k)}\right].
\end{align}
It is bounded from below by the entanglement of formation \cite{wootters1998entanglement} of the density operator, $E_f(\rho)$, and from above by the entropies of $\rho_A=\ptr{B}{\rho}$ and $\rho_B=\ptr{A}{\rho}$ 
\begin{align}\label{eq:EEbound}
E_f(\rho)\leq \overline E   \leq \min\left[S\lr{\rho_A},S\lr{\rho_B} \right].
\end{align}

The dependence of the EAEE on the unravelling becomes especially important  when the QT method is combined with MPS techniques to represent and propagate the (stochastic) wave function of the quantum many-body system. MPS representations are efficient as long as the bond dimension $\chi$, and thus the entanglement, is small. Since the computational cost of an MPS simulation grows exponentially with the entanglement, choosing the unravelling that minimizes the EAEE is of paramount importance. \textit{A priori} it is unclear which stochastic propagation scheme leads to small values of the EAEE, without constructing the trajectories~\cite{gharibian2008strong}. We can however predict how various types of the stochastic propagators affect the change rate of the EAEE at each instant of time and accordingly optimize it. This motivates a time-local greedy algorithm that continuously adapts the stochastic propagators to the conditional wave function as time progresses (Fig.~\ref{fig:Fig1}a). 

More specifically, given the pure state of the system $\ket{\phi^{(k)}(t)}$ of $k^{\rm th}$ trajectory at time $t$, we want to choose the stochastic propagator that minimizes the expected instantaneous entanglement increase rate, $\dot{\overline E} = d\overline E/dt$ (see Fig.~\ref{fig:Fig1}b). For simplicity we consider only propagators corresponding to independent measurements of each bath channel $j$ and jump operators that do not couple the partitions $A$ and $B$. As outlined above, the stochastic components of the propagator can then be applied in sequences. This allows to optimize the corresponding change rate of the EAEE, $\dot{\overline{E}}_j$, channel by channel.
Remarkably, one can analytically perform the minimization $\min_{X_j(t)}\left(\dot{\overline{E}}_j\right)$ over all stochastic propagators that correspond to measurements in eigenbases of arbitrary bath observables $X_j(t)=f[b_j(t), b_j^\dag (t)]$. Moreover, the minimum is always obtained either for a number measurement, $X_j(t)=b_j^\dag (t)b_j(t)$, or for a homodyne
measurement, $X_j(t) = b_j(t) e^{i\varphi_j} + b_j^\dagger(t) e^{-i\varphi_j}$. For the number measurement the EAEE change rate at a channel $j$ is
\begin{widetext}
\begin{align}
    \dot{\overline{E}}_j^\textrm{num} =  \tr{c_j\phi c_j^\dagger} \log_2\left[\tr{c_j\phi c_j^\dagger}\right] + \tr{\ptr{B}{c_j\phi c_j^\dagger} \left\{ \log_2\left[\ptr{B}{\phi}\right] - \log_2\left[\ptr{B}{c_j\phi c_j^\dagger}\right] \right\}}.
    \label{eq:E_num}
\end{align}
The measurement of the homodyne current gives instead:
\begin{align}
     \dot{\overline E}_j^\mathrm{hom}=\frac{1}{2\ln 2}\Bigg[\bigg|e^{-i\varphi_j}\tr{c_j\phi}+e^{i\varphi_j}\tr{\phi c_j^\dag}\bigg|^2 - \sum_{k, \ell}\frac{\ln\left(\xi_k\right)-\ln\left(\xi_\ell\right)}{\xi_k-\xi_\ell} \bigg|e^{-i\varphi_j}{\bra{\xi_k}\ptr{B}{c_j\phi}\ket{\xi_\ell}} + e^{i\varphi_j}{\bra{\xi_k}\ptr{B}{\phi c_j^\dag} \ket{\xi_\ell}} \bigg|^2\Bigg].\label{eq:E_Had}
\end{align}
\end{widetext}
Here $\phi=\ket{\phi_{j}^{(k)}(t)}\bra{\phi_{j}^{(k)}(t)}$ [see Eq.~\eqref{eq:propag}] and $\ptr{B}{\phi}\ket{\xi_\ell}=\xi_\ell\ket{\xi_\ell}$. We refer the reader to the Supplementary Materials (SM, Section III) for the proof and derivation of the above statements.

\begin{figure}[h]\centering
\includegraphics[width=\linewidth]{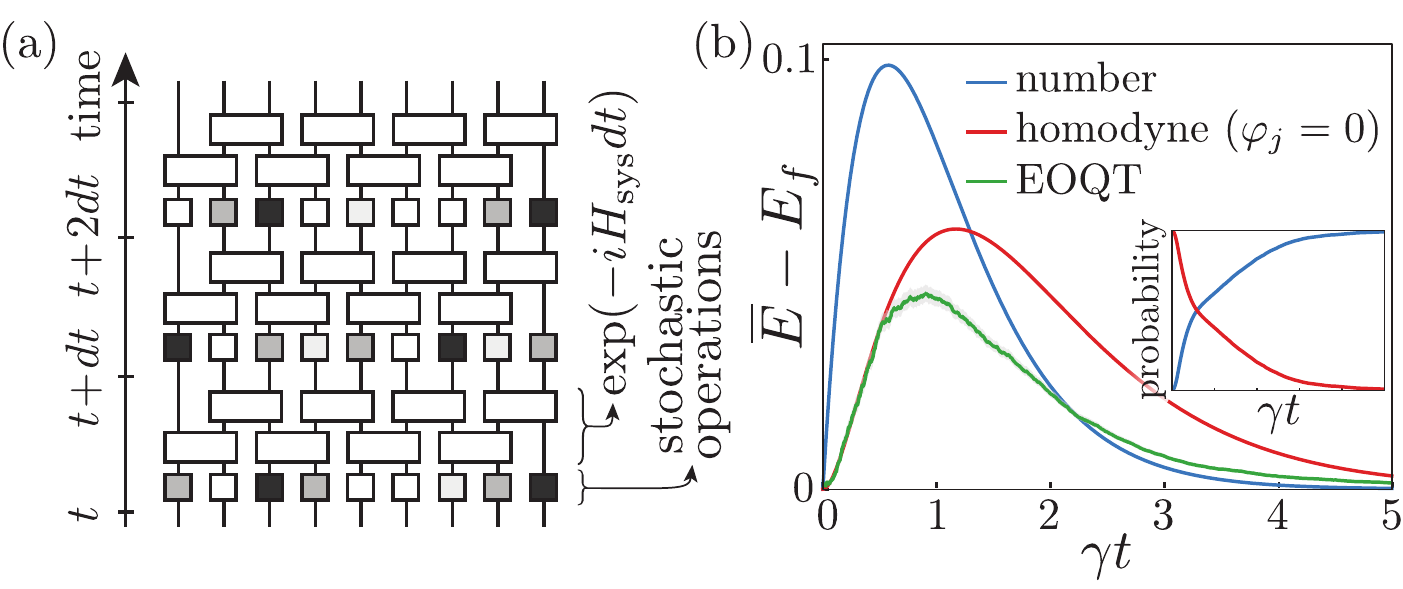}
\caption{(a) Schematic illustration of the EOQT-MPS algorithm. To propagate a state from $t$ to $t+dt$ we first apply a ``stochastic layer'': We calculate the unravelling that minimizes the expected EE in the next time step and apply the corresponding stochastic propagator. This procedure is implemented sequentially for all jump operators. Then we implement a ``deterministic layer'',  propagating the system with $H_{\rm sys}$ via standard methods such as TEBD. (b) Time dependence of the excess EAEE, $\overline{E}-E_f$, of a continuously monitored Bell pair for the number measurement (blue), the homodyne unravelling with $\varphi_j = 0$ (red), and the EOQT (green, number of trajectories $N = 10^4$). The inset shows the statistics of measurement choices of EOQT algorithm. The error bars are denoted by gray filling.
}
\label{fig:Fig2}
\end{figure}

This motivates the following entanglement-optimized quantum trajectory (EOQT) algorithm (see Fig.~\ref{fig:Fig2}a). For each discrete time step $dt$ propagate the state $\ket{\phi^{(k)}(t)}\rightarrow \ket{\phi^{(k)}(t+dt)}$ as follows:
\begin{enumerate}
    \item Sequentially for each channel $j$ calculate the EAEE change rates $\min_{\varphi_j} \dot{\overline{E}}_j^{\rm hom}$ and $\dot{\overline{E}}_j^{\rm num}$ and update the state according to the correspondingly optimal stochastic propagator;
    \item Complete the update of the resulting state by applying the coherent propagator $e^{-iH_{\rm sys}dt}$. 
\end{enumerate}
This EOQT algorithm can be naturally combined with MPS methods, such as the time evolving block decimation (TEBD) algorithm \cite{daleyTimedependentDensitymatrixRenormalizationgroup2004,whiteRealTimeEvolutionUsing2004a} (see Fig.~\ref{fig:Fig2}a). Importantly, the computational cost of the  evaluation and optimization of \eqref{eq:E_num} and \eqref{eq:E_Had} for an MPS with bond dimension $\chi$ is $\mathcal{O}(\chi^3d)$, where $d$ is the local Hilbert space dimension. This should be compared to the cost of the coherent propagation, which for each of the time steps $dt$ is $\mc{O}(\chi^3d^3 n)$. Thus the EAEE optimization does not significantly add to the cost of the standard propagation. On the other hand, the potential gain in the simulation efficiency can be substantial, as we show in the remainder of this paper. 

The simplest example illustrating the dependence of $\overline{E}$ on the unravelling is obtained by considering two qubits initially in a Bell state, $(\ket{00}+\ket{11}) / \sqrt{2}$, that are undriven ($H_{\rm sys}=0$) and coupled to a bath with a jump operator $c_j=\ket{1}_j\bra{1}$ and strength $\gamma_j=\gamma$ ($j=1,2$). In this case, $\overline{E}$ can be analytically calculated for different unravellings. For the photon number measurements, one obtains $\overline E ^\mathrm{num} = \sigma(2\gamma t)$, where we introduced the function:
\begin{align}
    \sigma(s)=\frac{1}{2\ln 2}\left[\left(1 + e^{-s}\right) \ln{\left(1 + e^{-s}\right)} + s e^{-s}\right].
\end{align}
The expression for homodyne measurement is
\begin{align}
    \overline E ^\mathrm{hom} (\tau) &=
    \frac{1}{2 \sqrt{2\pi\tau}}\int ds \sigma(s)e^{-\frac{(s-2\tau)^2}{8\tau}}
\end{align}
with $\tau=\gamma t \left(\cos^2\varphi_1+\cos^2\varphi_2\right)$. In Fig.~\ref{fig:Fig2} we compare the excess entanglements, $\overline E - E_f$, for number, homodyne, and optimal (EOQT) measurements. In this case $E_f$  evaluates to $
    E_f(t)=-r_+\log_2(r_+)-r_-\log_2(r_-),
$ where $r_\pm=\left(1\pm\sqrt{1-e^{-2\gamma t}}\right)/{2}$~\cite{wootters1998entanglement}. We note that for short times the homodyne unravelling with $\varphi_j=0$ ($j=1, 2$) saturates this bound, while for longer times the number unravelling approaches $E_f$ faster (Fig.~\ref{fig:Fig2}b).
The optimized algorithm indeed finds unravellings resulting in an EAEE that is always close to the theoretical minimum, $E_f$. As visible in the inset of Fig.~\ref{fig:Fig2}b, the optimizer chooses mostly the homodyne propagator with $\varphi_j=0$ for early times and switches over to predominantly choosing the number propagators around $\gamma t \approx 1$.  

To demonstrate the potential of our approach in a many-body setting, we consider a one-dimensional open random Brownian circuit (RBC) for a chain of spin-1/2 particles. We choose this example since the evolution under the RBC leads to rapid EE growth. The coherent part of the evolution is given by the time-dependent RBC Hamiltonian:
\begin{align}\label{eq:RBC}
   H_{\rm sys}(t) = \sum_{j=1}^{n-1}\sum_{k, \ell=0}^3 g_{j}^{k,\ell}(t) \sigma_j^k \otimes \sigma_{j+1}^\ell.
\end{align}
Here $n$ is the number of spins and $\sigma_j^k\in\{\mathbb{1}_j,\sigma_j^x,\sigma_j^y,\sigma_j^z\}$ are standard Pauli operators acting on spin $j$. The parameters $g_{j}^{k,\ell}(t)$ are Gaussian stochastic variables with $\smean{g_{j}^{k,\ell}(t)}=0$ and $\smean{g_{j}^{k,\ell}(t) g_{j'}^{k',\ell'}(t')}=\alpha\delta_{j,j'}\delta_{k,k'}\delta_{\ell,\ell'}\delta(t-t')$, where $\smean{\cdots}$ denotes the average over Hamiltonian realizations and $\alpha$ is the variance of RBC. The incohenert part is given by the jump operators $c_j=\sigma_j^z$ with uniform decay strength (measurement rate), $\gamma_j=\gamma$.

\begin{figure}[b]\centering
\includegraphics[width=\linewidth]{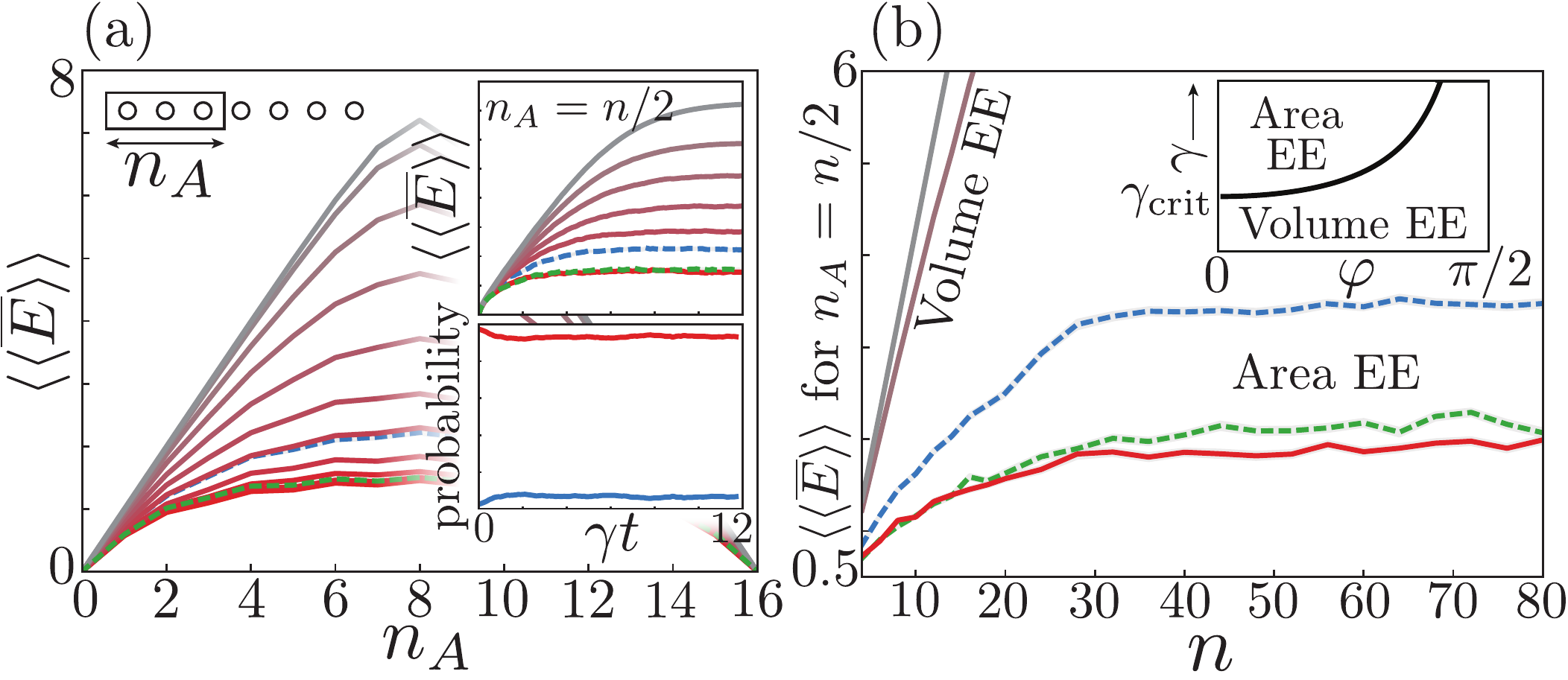}
\caption{EAEE for the open RBC. (a) EAEE profile in the long time limit for various unravellings. Solid lines denote homodyne unravellings with changing phase from $\varphi_j = 0$ (red) to $\varphi_j = \pi/2$ (grey) in increments of $\pi/20$, dashed blue and green lines correspond to the number and EOQT measurements. The insets show time evolution of EAEE for various unravellings and measurement choices of the EOQT. Here $m = n = 16$, $\chi = 128$. (b) Half-chain EAEE for larger systems (with $\chi = 512$). Depending on the homodyne phase, the ensembles display an area law at small $\varphi_j$ and a volume law at $\varphi_j\approx \pi/2$. The EOQT method results in an area-law EAEE close to the homodyne unravelling with $\varphi_j = 0$, whereas the number unravelling leads to an area law with larger EAEE. The phase diagram is shown in the inset. 
For all data in this figure we used $\alpha/\gamma = 0.1$ ($\alpha = 1$) and  $N = 200$. The statistical error bars lie within the lines' thickness.
}
\label{fig:Fig3}
\end{figure}

For sufficiently strong $\gamma$ this model exhibits a type of measurement-induced phase transition that is \textit{solely} driven by the type of unravelling (see Fig.~\ref{fig:Fig3}). Depending on the bath measurement operators, the character of EAEE changes from area law to volume law. That is, even though all types of bath measurements solve in the end the same ME~\eqref{eq:ME}, some unravellings lead to ensembles with area law EAEE that can be efficiently computed on classical computers, while other unravellings fail to do so.

This change from volume to area law can be shown explicitly by considering the stochastic propagator for a homodyne unravelling, $K_j$, defined above [see Eq.~\eqref{eq:propag_hom}]. Since ${\sigma_j^z}^\dag{\sigma_j^z}={\sigma_j^z}^2=\mathbb{1}$, it takes the form~\footnote{See Section~VI of the SM for more details.
}
\begin{align}
    K_j \propto \exp\left\{e^{i\varphi_j} \sqrt{\gamma}\sigma^z_j d\xi_j(t)\right\},
\end{align}
where $d\xi_j(t) = 2 \sqrt{\gamma} \left<\sigma^z_j\right>(t) \cos{\varphi_j} dt + dW_j(t)$. It has a unitary part, $\exp\left\{i\sin(\varphi_j) \sqrt{\gamma}\sigma^z_j d\xi_j(t) \right\}$, and a non-unitary part, $\exp\left\{\cos(\varphi_j) \sqrt{\gamma}\sigma^z_j d\xi_j(t)\right\}$. The unitary component can be absorbed into the coherent part of the evolution, as it leaves the ensemble of RBCs invariant. The non-unitary component can be re-interpreted a stochastic propagator of a Markov process with an effective decoherence rate $\gamma_j^{\rm eff}=\gamma\cos^2\varphi_j$. Changing the unravelling at fixed $\gamma$ by changing $\varphi_j$ is thus equivalent to changing $\gamma$ for a fixed unravelling. This equivalence is remarkable, as these two interpretations describe two profoundly different physical scenarios: The first refers to various representations of the solution of a single ME, while the second refers to particular solutions of various MEs. The latter has been studied extensively recently~\cite{liQuantumZenoEffect2018,liMeasurementdrivenEntanglementTransition2019,skinnerMeasurementInducedPhaseTransitions2019,chanUnitaryprojectiveEntanglementDynamics2019, gullansScalableProbesMeasurementinduced2020,choiQuantumErrorCorrection2020,bao2020theory,ippolitiEntanglementPhaseTransitions2021,zabaloCriticalPropertiesMeasurementinduced2020,zabaloOperatorScalingDimensions2021,botzungEngineeredDissipationInduced2021,mullerMeasurementinducedDarkState2021} and it is known that the conditional states can undergo a so-called measurement-induced phase transitions from area law to volume law entanglement, depending on the coupling strength to the environment. In our case, this phase transition occurs as a function of the unravelling, e.g.~parametrized by $\varphi_j$. Indeed, for $\varphi_j=\pi/2$ the QTs map to fully coherent RBC evolution, which generates entanglement that obeys a volume law. In the other limiting case, $\varphi_j=0$ (and sufficiently large $\gamma$), the measurement back-action continuously leads to an effective projection onto product states, resulting in entanglement within the system that satisfies an area law. These two phases are separated by a critical point at  $\gamma\cos^2\varphi_j = \gamma_{\rm crit}$. 
Thus, when $\gamma>\gamma_{\rm crit}$, the many-body ME can be efficiently solved using MPS via QTs \textit{if} an efficient unravelling is chosen. In Fig.~\ref{fig:Fig3} we show results of MPS simulations of various unravellings of the open RBC demonstrating that the EOQT method indeed finds efficiently computable unravelling. 

In summary, we introduced a novel QT method to simulate the time evolution of noisy quantum many-body system and showed that our method can enable an efficient classical simulation, where standard QT methods may fail. While in this work we focused on monitoring and optimizing each output channel separately, more general schemes, based on collective monitoring and optimization can be developed~\cite{holland1996measurement}. Detailed performance comparisons with other  techniques for open quantum many-body systems ~\cite{bonnes2014superoperators,cuiVariationalMatrixProduct2015a,wernerPositiveTensorNetwork2016,jinClusterMeanFieldApproach2016,whiteQuantumDynamicsThermalizing2018,biellaLinkedClusterExpansions2018,schwarzNonequilibriumSteadyStateTransport2018,hartmannNeuralNetworkApproachDissipative2019,vicentiniVariationalNeuralNetworkAnsatz2019,yoshiokaConstructingNeuralStationary2019,nagyVariationalQuantumMonte2019,weimerSimulationMethodsOpen2021} are left for future work.

\begin{acknowledgements}
We thank P. Zoller and B. Kraus for stimulating discussions. This work is supported by an ESQ Discovery Grant and an ERC Starting Grant (No. 101041435). 
\end{acknowledgements}

\bibliography{bibliography} 

\pagebreak
\widetext
\begin{center}
\textbf{\large Entanglement-Optimal Trajectories of Many-Body Quantum  Markov Processes: Supplementary material}
\end{center}
\setcounter{equation}{0}
\setcounter{figure}{0}
\setcounter{table}{0}
\setcounter{page}{1}
\makeatletter
\renewcommand{\theequation}{S\arabic{equation}}
\renewcommand{\thefigure}{S\arabic{figure}}
\renewcommand{\bibnumfmt}[1]{[S#1]}
\renewcommand{\citenumfont}[1]{S#1}

This Supplementary Material is structured as follows:
\begin{itemize}
    \item 
In Section~I we discuss adaptive quantum trajectory (QT) schemes on general grounds to confirm that such schemes indeed allow for an unbiased solution of master equations (MEs). 
\item
In Section~II and in Section~III we elaborate on \textbf{technical details underlying the main theoretical results} of our paper: We derive the the formulae for the ensemble-averaged entanglement entropy (EAEE) change rate for arbitrary environment monitoring schemes, obtain the expressions for number and homodyne measurement schemes, and prove the optimality result stated in the main text.  
\item
In Section~IV we provide details about the matrix product state (MPS) implementation of our entanglement-optimized quantum trajectory (EOQT) algorithm.
\item
In Section~V we apply our algorithm to several physically relevant examples of MEs that are chosen to illustrate, first, the convergence properties of the different QT schemes, and, second, the superior performance of the EOQT method compared to other schemes.
\item
Section~VI contains technical derivations underlying the discussion of the open random Brownian circuit (RBC) in the main text.
\end{itemize}
\section{Adaptive quantum trajectory methods} 
In this section we discuss general aspects regarding the validity of adaptive QT methods (such as the EOQT method introduced in the main text). While non-adaptive QT methods have been extensively discussed in the literature, adaptive schemes have received far less attention. For adaptive schemes it is sometimes not immediately obvious that no bias is introduced.  Our aim in this section is therefore to clarify that also adaptive continuous monitoring schemes give rise to valid trajectories, and thus that (properly constructed) adaptive QT methods lead to valid, unbiased solutions of the ME.  

We find it easiest to discuss this on very general grounds in the following. Let us consider a generic state $\ket{\Psi}$ of a system $S$ and its environment $E$. For the sake of this discussion, let us assume that the environment consists of a collection of several degrees of freedom, such that we can write  $E=e_1\otimes e_2\otimes e_3\otimes \dots$. Note that this includes the generic quantum-optical setting discussed in the main text, where each of these degrees of freedom is represented by a bosonic mode with creation and annihilation operators $b(t)$ and $b^\dag(t)$ (for notational simplicity we suppress the index $j$ here). The density operator $\rho$, describing the reduced state of the system $S$, is obtained by tracing over the degrees of freedom of $E$, $\rho=\textrm{tr}_{E}\left(\ket{\Psi}\bra{\Psi}\right)$. 
The notion of the density operator implies an inherent statistical interpretation: It describes the full measurement statistics an experimenter (called Alice) obtains from repeated preparations of the same state followed by measurements of observables in $S$, where any knowledge of $E$ is disregarded. 

Now, instead of disregarding $E$ for each such repetition, let us assume that another observer (called Bob) measures the degrees of freedom of the environment. For each repetition $i$, Bob obtains a set of measurement outcomes $\{x_{1}^{(i)},x_{2}^{(i)},\dots\}$, projecting the environment degrees of freedom onto the state $\ket{x_{1}^{(i)},x_{2}^{(i)},\dots}$. Here, $x_\ell^{(i)}$ denotes the measurement outcome of the $\ell$-th degree of freedom of $E$ in the $i$-th run of the experiment. Importantly, the probabilities for each outcome are determined by \emph{Born's rule}, such that $p^{(i)}\equiv p(x_1^{(i)},x_2^{(i)}\dots)=\textrm{tr}(\braket{x_1^{(i)},x_2^{(i)},\dots}{\Psi}\braket{\Psi}{{x_1^{(i)},x_2^{(i)},\dots}})$. Thus, Bob obtains a different description of the state $S$ for each run, namely, the state of $S$ conditional on the measurement outcomes of the environment: 
\begin{align}
    \ket{\psi^{(i)}}\equiv \ket{\psi(x_1^{(i)},x_2^{(i)}\dots)}\equiv \frac{\braket{x_{1}^{(i)},x_{2}^{(i)},\dots}{\Psi}}{p(x_1^{(i)},x_2^{(i)}\dots)^{1/2}}.
\end{align}
At first glance, this is in conflict with the description of the state of the system $S$ by Alice which assigns the density operator $\rho$ to every single repetition. The conflict is of course resolved by noting that the measurement of the environment precisely results in samples $\ket{\psi^{(i)}}$ that are distributed according to the probability $p^{(i)}$, which form a decomposition of $\rho=\sum_i p^{(i)}\ket{\psi^{(i)}}\bra{\psi^{(i)}}$. Thus, upon averaging over repetitions, Bob  obtains the same description of the state  as Alice. Mathematically, this is guaranteed by Born's rule: $$\rho=\ptr{E}{\ket{\Psi}\bra{\Psi}}=\sum_{x_1,x_2,\dots}\braket{x_1,x_2,\dots}{\Psi}\braket{\Psi}{x_1,x_2,\dots}=\sum_{x_1,x_2,\dots}p(x_1,x_2,\dots)\ket{\psi(x_1,x_2,\dots)}\bra{\psi(x_1,x_2,\dots)}.$$
This is the core idea behind QT methods, which numerically simulate the measurements of the environment by randomly drawing the measurement outcomes $\{x_1^{(i)},x_2^{(i)}\dots\}$ according to Born's probability distribution, thus obtaining samples $\ket{\psi^{(i)}}$ according to the distribution $p^{(i)}$.

A subtle but crucial point in the above discussion is that the description of the state of $S$ by Alice cannot (and does not) depend on the way that Bob measures the environment degrees of freedom. In particular, Bob can apply arbitrarily involved measurement strategies, including adaptive ones: While Bob's conditional states will depend on his measurement strategy, the ensemble of states he obtains \emph{due to Born's rule} is by construction always a valid decomposition of the reduced density operator. In fact, if this was not the case, Bob could send a signal to Alice by simply measuring the environment degrees of freedom, which is not possible by the basic principles of quantum mechanics.

Mathematically, we can show this as follows. Consider the following generic, adaptive, strategy for Bob: Instead of measuring all the environment degrees of freedom in a fixed basis, he chooses to first measure the first degree of freedom, $e_1$, in a given basis, spanned by states $\ket{x_1}$ (where $x_1=\{1,2,\dots \dim(e_1)\}$ labels basis states and potential measurement outcomes). The measurement outcome of this first measurement is random and distributed according to Born's rule, $p(x_1)=\tr{\braket{x_1}{\Psi}\braket{\Psi}{x_1}}$. Bob then chooses to measure the second degree of freedom, $e_2$, in a basis that depends on the measurement outcome of the measurement, $e_1$. We label these basis states by $\ket{x_2(x_1)}$ (where $x_2=\{1,\dots \dim(e_2)\}$). This notation explicitly indicates that the measurement basis for $e_2$ is determined by the measurement result of $e_1$, according to an \emph{adaptive} strategy specified by Bob. Of course, the outcome of $e_2$ is again random and distributed according to the probability distribution that can be obtained from Born's rule: $$p(x_2|x_1)=\tr{\braket{x_1,x_2(x_1)}{\Psi}\braket{\Psi}{x_1,x_2(x_1)}}/p(x_1)=p(x_1,x_2)/p(x_1).$$
Bob then continues and measures sequentially each environment degree of freedom, $e_j$, in a basis that he chooses adaptively, based on previous measurement outcomes, i.e. a basis spanned by states $\ket{x_j(x_{j-1},\dots, x_2,x_1)}$ (where $x_j\in \{1,2,\dots \dim(e_j)\}$, and we again highlight the dependence of the basis on previous results). For each $e_j$, Born's rule dictates the probability distribution of the measurement outcomes. By measuring all $n$ environment degrees of freedom, Bob obtains the state of $S$ conditional on the sequence of measurement outcomes:
\begin{align}
\frac{\braket{x_1,x_2(x_1),x_3(x_2,x_1),\dots x_n(x_{n-1},\dots, x_2,x_1)}{\Psi}}{\sqrt{p(x_1,x_2,x_3,\dots, x_n)}}
\end{align}
with a probability given by Born's rule as
\begin{align}p(x_1,x_2,x_3,\dots, x_n)=\mathrm{tr}\left[{\braket{x_1,x_2(x_1),\dots x_n(x_{n-1},\dots, x_2,x_1)}{\Psi}\braket{\Psi}{x_1,\dots x_n(x_{n-1},\dots, x_2,x_1)}}\right].\end{align}
We stress that the above expressions describe the adaptive measurement strategy of Bob, which includes the one considered in our work.
We can now easily show that the ensemble of states that Bob obtains forms a valid decomposition of $\rho$: 
\begin{align}
&\sum_{x_1,\dots,x_n}p(x_1,x_2,x_3,\dots, x_n)\frac{\braket{x_1,x_2(x_1),\dots x_n(x_{n-1},\dots,x_1)}{\Psi}\braket{\Psi}{x_1,x_2(x_1),\dots x_n(x_{n-1},\dots,x_1)}}{p(x_1,x_2,x_3,\dots, x_n)}\\
&=\sum_{x_1,\dots,x_{n-1}}\sum_{x_n}\braket{x_1,x_2(x_1),\dots x_n(x_{n-1},\dots,x_1)}{\Psi}\braket{\Psi}{x_1,x_2(x_1),\dots x_n(x_{n-1},\dots,x_1)}\\
&=\sum_{x_1,\dots,x_{n-1}}\mathrm{tr}_{{e_n}}\left[{\braket{x_1,x_2(x_1),\dots x_{n-1}(x_{n-2},\dots,x_1)}{\Psi}\braket{\Psi}{x_1,x_2(x_1),\dots x_{n-1}(x_{n-2},\dots,x_1)}}\right]\\
&=\sum_{x_1,\dots,x_{n-2}}\mathrm{tr}_{{e_{n-1}, e_n}}\left[{\braket{x_1,x_2(x_1),\dots x_{n-2}(x_{n-3},\dots,x_1)}{\Psi}\braket{\Psi}{x_1,x_2(x_1),\dots x_{n-2}(x_{n-3},\dots,x_1)}}\right]\\
&=\dots=\ptr{e_1,\dots, e_{n-1},e_n}{\ket{\Psi}\bra{\Psi}}\equiv \ptr{E}{\ket{\Psi}\bra{\Psi}}\equiv\rho.
\end{align}
Here we used the fact that the trace over the degrees of freedom can be taken in an arbitrary basis: $\ptr{e_j}{O}=\sum_{x_j}\bra{x_{j}(x_{j-1},\dots x_1)}O\ket{x_{j}(x_{j-1},\dots x_1)}$, which holds for any $\left(x_{j-1},\dots x_1\right)$, since the states $\bra{x_{j}(x_{j-1},\dots x_1)}$ form a orthonormal basis for every  $(x_{j-1},\dots x_1)$. 
In consequence, simulating such an adaptive measurement scheme to obtain QTs is equivalent to sampling pure states from a valid decomposition of the density operator (as long as the simulated measurements respect Born's rule).

\subsection{Toy example}
\begin{figure}[htb]\centering
\includegraphics[width=0.4\linewidth]{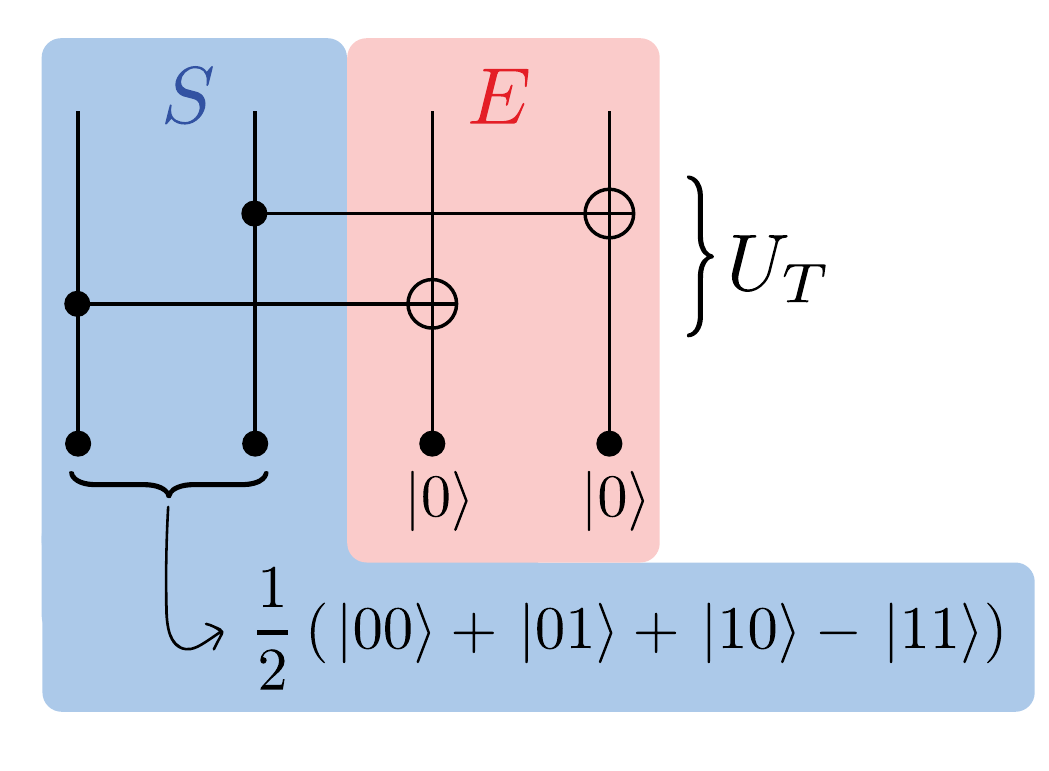}
\caption{Illustration of a toy example.
}
\label{fig:SM_Fig1}
\end{figure}

The above discussion is perhaps best illustrated by the following toy example. Consider a quantum system $S$ that consists of two qubits initially in the maximally entangled state $\ket{\psi_S} = \frac{1}{2}\left(\ket{00} + \ket{01} + \ket{10} - \ket{11}\right)$ and the environment $E$ consisting of two qubits in a product state $\ket{00}$ (see Fig.~\ref{fig:SM_Fig1}).

Imagine that two consecutive CNOT-operations $U_T$ were applied to first and second pair of system + environment qubits, as depicted in Fig.~\ref{fig:SM_Fig1}. These operations can be considered as elementary joint evolution of $S + E$. One can check that the state of $S$ after the evolution is a maximally mixed state:
\begin{align}
    \rho = \ptr{E}{U_T \ket{\psi_S} \bra{\psi_S} \otimes \ket{00} \bra{00} U_T^\dag} = \mathbb{1}/4.
    \label{eq:maxmix}
\end{align}
Now let us imagine that Bob measures the environment in some basis $\ket{x_i x_j}$ according to one of the three following strategies.

\subsubsection{First strategy}

Let us first consider Bob uses the computational basis, i.e.  $\{x_i x_j\} = \left\{00, 01, 10, 11\right\}$. This may be interpreted as a minimal model for a number measurement in the QT language. Depending on the measurement outcomes the conditional state of $S$ is
\begin{align}
    \ket{\psi_{00}} = \ket{00} \textrm{ with probability } p_{00} = {1}/{4}; \\
    \ket{\psi_{01}} = \ket{01} \textrm{ with probability } p_{01} = {1}/{4}; \\
    \ket{\psi_{10}} = \ket{10} \textrm{ with probability } p_{10} = {1}/{4}; \\
    \ket{\psi_{11}} = \ket{11} \textrm{ with probability } p_{11} = {1}/{4}.
\end{align}
In this case each of the measurement outcomes is a non-entangled product state. Summing the pure-state projectors with corresponding weights (probabilities) gives, as expected, the density matrix of the system after the evolution [as in Eq.~\eqref{eq:maxmix}]:
\begin{align}
    \frac{1}{4}\bigg(\ket{00}\bra{00} + \ket{01}\bra{01} + \ket{10}\bra{10} + \ket{11}\bra{11}\bigg) = \mathbb{1}/4.
\end{align}
\subsubsection{Second strategy}
Now let us consider instead Bob measuring the $E$ qubits in the states $\ket{\pm}=\frac{1}{\sqrt{2}}(\ket{0}\pm\ket{1})$, which gives the following span of measurement outcomes: $\{x_i x_j\} = \left\{++, +-, -+, --\right\}$. This may be considered as a minimal model analogous to a homodyne measurement in the QT language.  For each of these measurement outcomes, one can also find the corresponding conditional states
\begin{align}
    \ket{\psi_{++}} = \frac{1}{4} \bigg(\ket{00} + \ket{01} + \ket{10} - \ket{11}\bigg) \textrm{ with probability } p_{++} = {1}/{4}; \\
    \ket{\psi_{+-}} = \frac{1}{4} \bigg(\ket{00} - \ket{01} + \ket{10} + \ket{11}\bigg) \textrm{ with probability } p_{+-} = {1}/{4}; \\
    \ket{\psi_{-+}} = \frac{1}{4} \bigg(\ket{00} + \ket{01} - \ket{10} + \ket{11}\bigg) \textrm{ with probability } p_{-+} = {1}/{4}; \\
    \ket{\psi_{--}} = \frac{1}{4} \bigg(\ket{00} - \ket{01} - \ket{10} - \ket{11}\bigg) \textrm{ with probability } p_{--} = {1}/{4}.
\end{align}
As one can see, each of the measurement outcomes is a maximally-entangled state. Again, the  weighted sum gives the the expected maximally mixed state:
\begin{align}
    \frac{1}{4}\bigg(\ket{\psi_{++}}\bra{\psi_{++}} + \ket{\psi_{+-}}\bra{\psi_{+-}} + \ket{\psi_{-+}}\bra{\psi_{-+}} + \ket{\psi_{--}}\bra{\psi_{--}}\bigg) = \mathbb{1}/4.
\end{align}
\subsubsection{Third strategy}
Now let us turn to a simple example of \textit{adaptive measurement}. Imagine that Bob first measures the first environment qubit and, depending on the outcome, chooses the measurement basis for the second qubit. For example, if the first qubit, measured in the computational basis, results in a outcome $0$, then the second qubit is also measured in the computational basis. On the other hand, if the first qubit gives $1$ as an outcome, the second qubit should be measured in ``$\pm$''-basis. This situation produces the following span of outcomes: $\{x_i x_j\} = \left\{00, 01, 1+, 1-\right\}$. As before, we can find the corresponding trajectories:
\begin{align}
    &\ket{\psi_{00}} = \frac{1}{2}\ket{00} \textrm{ with probability } p_{00} = {1}/{4}; \\
    &\ket{\psi_{01}} =  \frac{1}{2} \ket{01} \textrm{ with probability } p_{01} = {1}/{4}; \\
    &\ket{\psi_{1+}} = \frac{1}{2 \sqrt{2}} \bigg(\ket{10} - \ket{11}\bigg) \textrm{ with probability } p_{1+} = {1}/{4}; \\
    &\ket{\psi_{1-}} = \frac{1}{2 \sqrt{2}} \bigg(\ket{10} + \ket{11}\bigg) \textrm{ with probability } p_{1-} = {1}/{4}.
\end{align}
And, as before, this adaptive measurement gives the expected state of the system, the maximally-mixed density matrix:
\begin{align}
    \frac{1}{4}\bigg(\ket{\psi_{00}}\bra{\psi_{00}} + \ket{\psi_{01}}\bra{\psi_{01}} + \ket{\psi_{1+}}\bra{\psi_{1+}} + \ket{\psi_{1-}}\bra{\psi_{1-}}\bigg) = \mathbb{1}/4.
\end{align}
As one can see, independent of the measurement strategy, Bob always obtains pure samples (trajectories) from various decompositions of  the same density operator. This example further illustrates that different measurement strategies lead to trajectories with different entanglement properties. 

\section{Entanglement of conditional ensembles} 
\label{sec:1}
In this section we give a general expression of the average entanglement entropy of an ensemble of conditional states, which will allow us to calculate the EAEE change rate in QT in the next section. This forms the basis of one of the main results of our work. 

We consider a closed tripartite system $A \otimes B\otimes M$. The state of the full system $A\otimes B\otimes M$ can be written as
\begin{align}\ket{\Psi}=\sum_{\vec{x}}  \ket{\tilde\phi(\vec{x})}\ket{\vec{x}},
\end{align}
where $\ket{\tilde\phi(\vec{x})}$ is the unnormalized state of $A$ and $B$, conditional on $M$ being in state $\ket{\vec{x}}$. Here the states $\vec{x}$ form an orthonormal basis of $M$, $\braket{\vec{x}}{\vec{y}}=\delta_{\vec{x},\vec{y}}$. That is, if $M$ is measured in this basis, the system $A\cup B$ is projected onto the state 
\begin{align}\label{eq:ensemble}
\ket{\tilde\phi(\vec{x})}=\braket{\vec{x}}{\Psi} \textrm{ with probability } p(\vec{x})=\braket{\tilde\phi(\vec{x})}{\tilde\phi(\vec{x})}.
\end{align}
It is convenient to introduce normalized states $\ket{\phi(\vec{x})}=\ket{\tilde\phi(\vec{x})}/\sqrt{p(\vec{x})}$, and the notation $\Psi=\ket{\Psi}\bra{\Psi}$,  $\tilde\phi(\vec{x}) =\ket{\tilde\phi(\vec{x})}\bra{\tilde\phi(\vec{x})}$ and $\phi(\vec{x})=\ket{\phi(\vec{x})}\bra{\phi(\vec{x})}$. Importantly, the ensemble of states \eqref{eq:ensemble} depends on the choice of basis in which $M$ is measured.
Of course the density operator of $A\cup B$, i.e. the ensemble average of the conditional states $\phi(\vec{x})$, is manifestly invariant under changes of the measurement basis:
\begin{align}
\rho=\sum_{\vec{x}}p(\vec{x})\phi(\vec{x})=\sum_{\vec{x}}\tilde\phi(\vec{x})=\ptr{{M}}{\Psi},
\end{align}
However, the ensemble average of any non-linear functional of the conditional states in general does depend on the basis in which $M$ is measured.

We are interested in the entanglement properties of the ensemble of states \eqref{eq:ensemble}, specifically the EAEE of the bipartite splitting of $A$ and $B$:
\begin{align}
\overline E = \sum_{\vec{x}}p(\v{x}) E\left[\phi(\v{x})\right].
\end{align}
To proceed, it is convenient to define the operator 
\begin{align}
\mu=\sum_{\v{x}}\tilde\phi(\vec{x})\otimes\ket{\vec{x}}\bra{\vec{x}}=\sum_{\v{x}}p(\vec{x})\phi(\vec{x})\otimes\ket{\vec{x}}\bra{\vec{x}}.
\end{align}
Note that $\mu$ is dependent on the specific choice of basis in $M$. We define $\mu_{M}=\ptr{A,B}{\mu}=\sum_{\v{x}}p(\vec{x})\ket{\vec{x}}\bra{\vec{x}}$,  $\mu_{A,M}=\ptr{B}{\mu}=\sum_{\v{x}}p(\vec{x})\rho_A(\vec{x})\otimes\ket{\vec{x}}\bra{\vec{x}}$, and $\rho_A(\vec{x})=\ptr{B}{\phi(\vec{x})}$. With this we define the \textit{conditional entropy} of the state $\mu$ as
\begin{align}
S(A|M)&=S(\mu_{AM})-S(\mu_M),
\end{align}
where $S(X) = -\mathrm{tr}\left[X \log_2 \left(X\right)\right]$ is the von Neumann entropy. The conditional entropy is an important quantity since it is precisely equivalent to the EAEE of the ensemble \eqref{eq:ensemble}:  
\begin{align}
S(A|M)&=-\mathrm{tr}\left[{\sum_{\v{x}}p(\vec{x})\rho_A(\vec{x})\otimes\ket{\vec{x}}\bra{\vec{x}}\log_2 \sum_{\v{y}}p(\vec{y})\rho_A(\vec{y})\otimes\ket{\vec{y}}\bra{\vec{y}}}\right]+\mathrm{tr}\left[\sum_{\v{x}}p(\vec{x})\ket{\vec{x}}\bra{\vec{x}}\log_2 \sum_{\v{y}}p(\vec{y})\ket{\vec{y}}\bra{\vec{y}}\right]\\
&=-\sum_{\v{x}}p(\vec{x})\mathrm{tr}\left\{{\rho_A(\vec{x})\log_2 \left[p(\vec{x})\rho_A(\vec{x})\right]}\right\}+\sum_{\v{x}}p(\vec{x})\log_2 p(\vec{x})\\
&=-\sum_{\v{x}}p(\vec{x})\left\{{\mathrm{tr}\left[{\rho_A(\vec{x})\log_2\rho_A(\vec{x})}\right]+\mathrm{tr}\left[{\rho_A(\vec{x})\log_2 p(\vec{x})}\right]-\log_2 p(\vec{x})}\right\}\\
&=-\sum_{\v{x}}p(\vec{x})\mathrm{tr}\left[{\rho_A(\vec{x})\log_2\rho_A(\vec{x})}\right]=\sum_{\vec{x}}p(\v{x}) S\left[\rho_A(\v{x})\right] =\sum_{\vec{x}}p(\v{x}) E\left[\phi(\v{x})\right] \equiv \overline E.\label{eq:CE}
\end{align}
Note that, since $S(\rho_A)=S(\rho_B)$ for a pure state on $A\cup B$, we have $S(A|M)=S(B|M)$. 
In the following sections we apply this to derive expressions for the change rate of the EAEE in a QT simulation.

\section{Entanglement entropy change rates in Monitored systems }
\subsection{Monitored systems}
\label{susec:monsys}
 We are interested in an open quantum many-body system with continuously measured output ports, described by the Hamiltonian 
\begin{align}\label{eq:SMH}
H(t)=H_{\rm sys} + i \sum_{j} \sqrt{\gamma_j}\left[b_j(t)^\dag c_j-c_j^\dag b_j(t)\right] = H_{\rm sys} + i \sum_{j} H_j(t)
\end{align}
with $[b_j(t),b_{j'}(t')^\dag]=\delta_{j,j'}\delta(t-t')$ and $b_j(t)\vac=0~\forall$\textit{j}. Here the quantum many-body system takes role of the $A\cup B$ in Section~\ref{sec:1}, and the bosonic modes of the environment play the role of $M$. We are specifically interested in the situation, where we are given a pure state of $A\cup B$, $\ket{\phi(t)}$, propagate the the state of system and environment $\ket{\Psi(t)}=\ket{\phi(t)}\vac$ by a small time step $dt$ using the Hamiltonian \eqref{eq:SMH}, to obtain the state $\ket{\Psi(t+dt)}$, and then measure the state of the modes of the environment (separately for each channel), to obtain a state $\ket{\phi(t+dt)}$, of $A\cup B$, which depends on the measurement basis and measurement outcome. As detailed in the main text, this procedure can be decomposed in a coherent component and a sequence of stochastic components. 
Specifically, one defines the sequence of states $\ket{\phi_{j}}$ as follows: Staring form the state $\ket{\phi_{1}}=\ket{\phi(t)}$, we sequentially (for $j=1,\dots m$) do the following:
\begin{enumerate}
\item Propagate the state $\ket{\phi_{j}}\vac$ of the full system $A\cup B\cup M$ with the unitary 
\begin{align}
U_j = \mc{T}\exp\left[\int_{t}^{t+dt} dt' H_j(t')\right] \cong \exp\left[\lambda(b_j^\dag c_j-c_j^\dag b_j)\right],
\end{align}
where the last equation holds in the limit of small $dt$ and we introduced the normalized quantum Ito increment $b_j = \frac{1}{\sqrt{dt}}\int_{t}^{t+dt} dt' b_j(t)$, (satisfying $[b_j, b_j^\dag]=1$ and $b_j\vac=0$). We also introduced $\lambda=\sqrt{dt}$ for notation simplicity in what follows.
This gives the state $\ket{\Psi_{j}}=U_j\ket{\phi_j}\vac$;
\item Measure the state of the mode $b_j$ in an arbitrary orthonomal basis, specified by basis vectors $\ket{x}$. Given the measurement outcome $x$, we obtain the state of $A\cup B$ as $\ket{\phi_{j+1}}=\braket{x}{\Psi_j}$.
\end{enumerate}
Finally, we obtain the state $\ket{\phi(t+dt)}=\exp(-iH_{\rm sys}dt)\ket{\phi_{m+1}}$.
The EE changes during each of the steps $\ket{\phi_j}\rightarrow \ket{\phi_{j+1}}$, as well as in the last, coherent step.  While the coherent step is fixed, the steps $\ket{\phi_j}\rightarrow \ket{\phi_{j+1}}$ depend on the choice of the measurement basis. We are thus interested in the average change rate of the EE in these steps $\dot{\overline{E}}_j=\lim_{dt\rightarrow 0}\frac{1}{dt}(\overline{E(\phi_{j+1})}-E(\phi_j))$, where the line indicates the average over measurement outcomes.

\subsection{Entanglement entropy change rate}
\label{susec:eecr}
To calculate $\dot{\overline{E}}_j$, we apply the formalism developed in Section~\ref{sec:1} to the state $\ket{\Psi_j}$. We identify the measured  bosonic mode (i.e. the mode with bath annihilation and creation operators $b_j$ and $b_j^\dag$) with $M$. To proceed we assume that the jump opearator $c_j$ is fully supported in either $A$ or $B$ (without loss of generality we consider the case when it's support is in $B$).  With this we identify $\overline{E(\phi_{j+1})}=S(A|M)=S(B|M)$, which can be obtained via \eqref{eq:CE}, using 
\begin{align}
\mu_{AM}&=\sum_{x}\bra{x} \ptr{B}{U_j\ket{\phi_j}\bra{\phi_j}\otimes \ket{0}\bra{0} U_j^\dag} \ket{x}\otimes \ket{x}\bra{x}\\
&=\sum_{x}\ptr{B}{\bra{x}U_j\ket{0}\ket{\phi_j}\bra{\phi_j} \bra{0}U_j^\dag \ket{x}}\otimes \ket{x}\bra{x}\\
&=\sum_{x}\ptr{B}{M_x\phi M_x^\dag}\otimes \ket{x}\bra{x},
\end{align}
where we defined $M_x=\bra{x}U_j\ket{0}$ and $\phi = \ket{\phi_j}\bra{\phi_j}$ (suppressing the  index $j$ from here on when clear from the context), and  $\ket{0}$ is the vacuum state, i.e.  $b_j\ket{0}=0$. We also have:
\begin{align}
\mu_{M}
&=\sum_{x}\tr{M_x\phi M_x^\dag} \ket{x}\bra{x}.
\end{align}
For the entropies we can write: 
\begin{align}
S(\mu_M)&=-\sum_x \tr{M_x\phi M_x^\dag}\log_2 \tr{M_x\phi M_x^\dag},\\
S(\mu_{AM})&=-\sum_x \mathrm{tr}\left[\ptr{B}{M_x\phi M_x^\dag}\log_2 \ptr{B}{M_x\phi M_x^\dag}\right].
\end{align}
This suggest to introduce the notation $\phi_x=M_x\phi M_x^\dag$. We remark that $M_x$ depends on $\lambda$.
\\
\\

In the following we calculate $\dot{\overline{E}}_j$ from a Taylor expansion of 
$\overline{E(\phi_{j+1})}$ up to second order in $\lambda$, i.e. $\overline{E(\phi_{j+1})}=S(A|M)|_{\lambda=0}+\lambda \partial\lambda S(A|M)|_{\lambda=0}+\frac{1}{2}\lambda^2 \partial^2_\lambda S(A|M)|_{\lambda=0}$. The explicit calculation of these terms below leads to $\partial_\lambda S(A|M)|_{\lambda=0}=0$, and together with $E(\phi_{j})=S(A|M)|_{\lambda=0}$ we find 
\begin{align}
\dot{\overline{E}}_j=\frac{1}{2}\partial^2_\lambda S(A|M)|_{\lambda=0}.
\end{align}
To evaluate the terms in this Taylor expansion we use $\partial_\lambda^k M_x=\bra{x}U(b^\dag c-c^\dag b)^k\ket{0}$. Since $U|_{\lambda=0}=1$, we have
\begin{align}
&M_x|_{\lambda=0}=\braket{x}{0},\\
&\partial_\lambda M_x|_{\lambda=0}=\braket{x}{1}c,\\
&\partial_\lambda^2 M_x|_{\lambda=0}=-\braket{x}{0}c^\dag c+\braket{x}{2}\sqrt{2}c^2,
\end{align}
where $\ket{n}=\frac{{b^\dag}^n}{\sqrt{n!}}\ket{0}$. We then find the derivatives of $\phi_x$:
\begin{align}
&\phi_x|_{\lambda=0}=|\braket{x}{0}|^2\phi,\\
&\partial_\lambda \phi_x|_{\lambda=0}=\braket{x}{1}\braket{0}{x}c\phi + \braket{x}{0}\braket{1}{x}\phi c^\dag,\\
&\partial_\lambda^2 \phi_x|_{\lambda=0}=
-|\braket{x}{0}|^2(c^\dag c\phi+\phi c^\dag c)+
2|\braket{x}{1}|^2c\phi c^\dag
+\sqrt{2}\braket{x}{2}\braket{0}{x}c^2\phi+\sqrt{2}\braket{x}{0}\braket{2}{x}\phi {c^\dag}^2. \end{align} 

It is useful to introduce the following:
\begin{align}
&\braket{x}{j} \rightarrow \zeta_j(x),\\
&\braket{x}{0} \rightarrow \zeta_0(x) = |\zeta_0(x)| e^{i\mu_0(x)},\label{notat1} \\
&\braket{x}{1} \rightarrow \zeta_1(x) = |\zeta_1(x)| e^{i\mu_1(x)},\label{notat2} \\
&\textrm{and~}\mu_{01}(x) = \mu_0(x) -\mu_1(x).\label{notat3}
\end{align}

\subsubsection{First derivative: $\partial_\lambda S(A|M)$}
With this we can calculate the derivatives of $S(\mu_M)$ and $S(\mu_{AM})$, which formally are obtained as 
\begin{align}
\partial_\lambda S(\mu_M)&=-\sum_x {\tr{\partial_\lambda \phi_x}\log_2 \tr{\phi_x}+\partial_\lambda\tr{\phi_x}/\ln 2}=-\sum_x \tr{\partial_\lambda \phi_x}\log_2 \tr{\phi_x}\\
\partial_\lambda S(\mu_{AM})&=-\sum_x {\mathrm{tr}\left[{\ptr{B}{\partial_\lambda \phi_x}\log_2 \ptr{B}{\phi_x}}\right]+\partial_\lambda\tr{\phi_x}/\ln 2}=-\sum_x \mathrm{tr}\left[{\ptr{B}{\partial_\lambda \phi_x}\log_2 \ptr{B}{\phi_x}}\right].
\end{align}
We can then evaluate those expressions at $\lambda=0$ and obtain:
\begin{align}
\partial_\lambda S(\mu_M)|_{\lambda=0}&=-\sum_x \mathrm{tr}\left({\zeta_1(x)\zeta^*_0(x)c\phi + \zeta_0(x)\zeta^*_1(x)\phi c^\dag}\right)\log_2 \tr{|\zeta_0(x)|^2\phi}\\
&=-\tr{c\phi} \sum_x \zeta_1(x)\zeta^*_0(x)\log_2|\zeta_0(x)|^2+\rm{c.c.}\\
\partial_\lambda S(\mu_{AM})|_{\lambda=0}&=-\sum_x \mathrm{tr}\left[{\ptr{B}{\zeta_1(x)\zeta^*_0(x)c\phi + \zeta_0(x)\zeta^*_1(x)\phi c^\dag}\log_2 \ptr{B}{|\zeta_0(x)|^2\phi}}\right]\\
&=\partial_\lambda S(\mu_{M})|_{\lambda=0}-\sum_x \mathrm{tr}\left[{\ptr{B}{\zeta_1(x)\zeta^*_0(x) c\phi + \zeta_0(x)\zeta^*_1(x)\phi c^\dag}\log\ptr{B}{ \phi}}\right]\\
&=\partial_\lambda S(\mu_{M})|_{\lambda=0}.
\end{align}
Here we used the property $\sum_x\braket{x}{j}\braket{k}{x}=\braket{k}{j}=\delta_{k, j}$. In summary we have 
\begin{align}
&\partial_\lambda S(A|M)|_{\lambda=0}=0,
\end{align}
as stated above.
\subsubsection{Second derivative: $\partial^2_\lambda S(A|M)$}
We now proceed to calculate 
\begin{align}
\partial_\lambda^2 S(\mu_M)&=-\sum_x \tr{\partial_\lambda^2 \phi_x}\log_2 \tr{\phi_x}+\frac{\left[\tr{\partial_\lambda \phi_x}\right]^2}{\tr{\phi_x}\ln 2},\\
\partial_\lambda^2 S(\mu_{AM})&=-\sum_x \mathrm{tr}\left[{\ptr{B}{\partial_\lambda^2 \phi_x}\log_2 \ptr{B}{\phi_x}} + \ptr{B}{\partial_\lambda \phi_x} \partial_\lambda \log_2\ptr{B}{\phi_x}\right].
\end{align}
Again, we are interested in evaluating this at $\lambda=0$. More precisely, we are interested in the quantity 
\begin{align}
\dot{\overline E}_j\equiv \frac{1}{2}\partial_\lambda^2S(A|M)|_{\lambda=0}=\frac{1}{2}\lim_{\lambda\rightarrow 0}&\left[\partial_\lambda^2 S(\mu_{AM})-\partial_\lambda^2 S(\mu_M)\right]\\
=\frac{1}{2\ln 2}\lim_{\lambda\rightarrow 0}\bigg\{&\sum_x \tr{\partial_\lambda^2 \phi_x}\ln \tr{\phi_x} - \mathrm{tr}\left[{\ptr{B}{\partial_\lambda^2 \phi_x}\ln \ptr{B}{\phi_x}}\right]\label{F_first}\\
&+ \frac{\left[\tr{\partial_\lambda \phi_x}\right]^2}{\tr{\phi_x}} - \mathrm{tr}\left[{\ptr{B}{\partial_\lambda \phi_x}R^{-1}_{\ptr{B}{\phi_x}}(\ptr{B}{\partial_\lambda \phi_x})}\right]\bigg\}.\label{F_last}
 \end{align}
Here we introduced the logarithmic derivative, defining  
 \begin{align}
 R^{-1}_{\rho}(X)=\sum_{j,k}\frac{\ln r_j - \ln r_j}{r_j-r_k}\bra{r_j}X\ket{r_k}\ket{r_j}\bra{r_k}
 \end{align}
 with $\rho\ket{r_k}=r_k\ket{r_k}$. 
 The above expression for the EAEE time-derivative $\dot{\overline E}_j$ is valid for an arbitrary measurement of the environement mode of the channel $j$. Note that evaluating the limit $\lambda\rightarrow 0$ requires special attention if the measurement basis contains a state $\ket{x}$ such that $\braket{x}{0} = 0$, which we refer to as \textit{vacuum-resolving measurements}, i.e. the measurements that exclude the vacuum state of the bath. If there is no such state, we name the measurement as \textit{non-vacuum-resolving}.  We consider these cases separately in the following.

\subsection{Non-vacuum-resolving measurement}\label{sss:nd}
Let us consider the case where $\zeta_0(x)\neq 0$ for all $x$. And let us look now at each term in the expression for  $\dot{\overline E}_j$, i.e. \eqref{F_first} and \eqref{F_last}, individually.
The first two terms in \eqref{F_first} can be written as
 \begin{align}
 \sum_x& \tr{\partial_\lambda^2 \phi_x}\ln \tr{\phi_x} - \tr{\ptr{B}{\partial_\lambda^2 \phi_x}\ln \ptr{B}{\phi_x}} \label{F_log}\\
 = \sum_x& \left(|\zeta_1(x)|^2 - |\zeta_0(x)|^2\right) \tr{2c\phi c^\dag}\ln{\left\{|\zeta_0(x)|^2 + \mathrm{tr}\left[{O(x,\lambda)}\right]\right\}} \\
 -\sum_x& \mathrm{tr} \left(\left[|\zeta_1(x)|^2 \ptr{B}{2 c\phi c^\dag} - |\zeta_0(x)|^2 \ptr{B}{c^\dag c \phi + \phi c^\dag c }\right] \ln{\left\{|\zeta_0(x)|^2 \ptr{B}{\phi} + \mathrm{tr}_\mathrm{B}\left[{O(x,\lambda)}\right]\right\}}\right), \end{align}
where $O(x,\lambda) = \lambda \zeta_1(x) \zeta_0^*(x)  c \phi + \textrm{h.c.}+\mathcal{O}(\lambda^2)$, which is not relevant in the limit $\lambda\rightarrow 0$ in the non-vacuum-resolving case. In this limit we get
 \begin{align}
 \lim_{\lambda \rightarrow 0} &\sum_x \tr{\partial_\lambda^2 \phi_x}\ln \tr{\phi_x} - \tr{\ptr{B}{\partial_\lambda^2 \phi_x}\ln \ptr{B}{\phi_x}} = \sum_x \left(|\zeta_1(x)|^2 - |\zeta_0(x)|^2\right) \tr{2c\phi c^\dag}\ln{\left[|\zeta_0(x)|^2\right]} \\
 &-\sum_x \mathrm{tr} \left\{\left[|\zeta_1(x)|^2 \ptr{B}{2 c\phi c^\dag} - |\zeta_0(x)|^2 \ptr{B}{c^\dag c \phi + \phi c^\dag c }\right] \ln{\left[|\zeta_0(x)|^2 \ptr{B}{\phi}\right]}\right\}\\
 =
 &-\sum_x \mathrm{tr} \left\{\left(|\zeta_1(x)|^2  - |\zeta_0(x)|^2 \right)\ptr{B}{2 c\phi c^\dag} \ln{\left[ \ptr{B}{\phi}\right]}\right\}=0.
 \end{align}
 In the last line we first made use of the fact that $c$ acts non-trivially only in $B$, which allows us to use the cyclic property of the partial trace, and then we used  $\sum_x\braket{x}{j}\braket{k}{x}=\braket{k}{j}=\delta_{k, j}$ [we remind the reader of the short-hand notations from Eqs.~(\ref{notat1}) and (\ref{notat2})].

The first term in  \eqref{F_last} can be expressed as follows:
\begin{align}
    \sum_x\frac{1}{2\ln 2}\lim_{\lambda \rightarrow 0}\frac{\left[\tr{\partial_\lambda \phi_x}\right]^2}{\tr{\phi_x}}=\sum_x\frac{|\zeta_1(x)|^2}{2\ln 2}\bigg|e^{-i\mu_{01}(x)}\tr{c\phi} + e^{+i\mu_{01}(x)}\tr{\phi c^\dag}\bigg|^2.
\end{align}

For the last term in \eqref{F_last} we first note that 
  \begin{align}
 \lim_{\lambda \rightarrow 0}R^{-1}_{\ptr{B}{\phi_x}}(X)|_{\lambda=0}= R^{-1}_{|\zeta_0(x)|^2\ptr{B}{\phi}}(X)= \frac{1}{|\zeta_0(x)|^2}R^{-1}_{\ptr{B}{\phi}}(X)=\frac{1}{|\zeta_0(x)|^2}\sum_{j,k}\frac{\ln \xi_j- \ln\xi_k}{\xi_j-\xi_k}\bra{\xi_j}X\ket{\xi_k}\ket{\xi_j}\bra{\xi_k}
 \end{align}
  with $\ptr{B}{\phi}\ket{\xi_k}=\xi_k\ket{\xi_k}$, {i.e.} $\xi_k$'s and $\ket{\xi_k}$'s are the eigenvalues and the eigenvectors of the reduced state of subsystem $A$, correspondingly. With this we write the last term as 
  \begin{align}
     -\frac{1}{2\ln 2}\sum_x |\zeta_1(x)|^2 \sum_{j,k}\frac{\ln\xi_j - \ln\xi_k}{\xi_j - \xi_k} \bigg|e^{-i\mu_{01}(x)} \bra{\xi_j}\ptr{B}{c\phi}\ket{\xi_k} + e^{+i\mu_{01}(x)}\bra{\xi_j}\ptr{B}{\phi c^\dag}\ket{\xi_k}\bigg|^2.\\
  \end{align}
In summary, we obtain for non-vacuum-resolving measurement the following expression for the time-derivative of the EAEE:
\begin{align}
    \dot{\overline E}_\mathrm{NVR}= \frac{1}{2\ln 2}\sum_x |\zeta_1(x)|^2 \Bigg[&\bigg|e^{-i\mu_{01}(x)}\tr{c\phi} + e^{+i\mu_{01}(x)}\tr{\phi c^\dag}\bigg|^2 -  \label{F_non-degen1}\\
    &
    - \sum_{j,k}\frac{\ln\xi_j - \ln\xi_k}{\xi_j - \xi_k}\bigg|e^{-i\mu_{01}(x)}\bra{\xi_j}\ptr{B}{c\phi}\ket{\xi_k} + e^{+i\mu_{01}(x)}\bra{\xi_j}\ptr{B}{\phi c^\dag}\ket{\xi_k}\bigg|^2\Bigg]. \label{F_non-degen2}
\end{align}
Both terms exhibit a dependence on the measurement basis, specifically on its ``phase'' $\mu_{01}(x)$.

\subsection{Vacuum-resolving measurement}
In the vacuum-resolving case we need to consider the terms in the sum over $x$ where $\zeta_0(x) = 0$  separately. Thus we split the sum in $\dot{\overline E}_j$ into two parts, one part with $\zeta_0(x) \neq 0$ and another part with $\zeta_0(x) = 0$:
\begin{align}
\dot{\overline E}_j
=&\frac{1}{2\ln 2}\lim_{\lambda\rightarrow 0}\bigg(\sum_{x|\zeta_0(x)\neq 0} \tr{\partial_\lambda^2 \phi_x}\ln \tr{\phi_x} - \mathrm{tr}\left[{\ptr{B}{\partial_\lambda^2 \phi_x}\ln \ptr{B}{\phi_x}}\right]\label{F1_deg}\\
&+ \frac{\left[\tr{\partial_\lambda \phi_x}\right]^2}{\tr{\phi_x}} - \mathrm{tr}\left\{{\ptr{B}{\partial_\lambda \phi_x}R^{-1}_{\ptr{B}{\phi_x}}\left[\ptr{B}{\partial_\lambda \phi_x}\right]}\right\}\bigg)\label{F2_deg}\\
+&\frac{1}{2\ln 2}\lim_{\lambda\rightarrow 0}\bigg(\sum_{x|\zeta_0(x) = 0} \tr{\partial_\lambda^2 \phi_x}\ln \tr{\phi_x} - \mathrm{tr}\left[{\ptr{B}{\partial_\lambda^2 \phi_x}\ln \ptr{B}{\phi_x}}\right]\label{F3_deg}\\
&+ \frac{\left[\tr{\partial_\lambda \phi_x}\right]^2}{\tr{\phi_x}} - \mathrm{tr}\left\{{\ptr{B}{\partial_\lambda \phi_x}R^{-1}_{\ptr{B}{\phi_x}}\left[\ptr{B}{\partial_\lambda \phi_x}\right]}\right\}\bigg).\label{F4_deg}
 \end{align}
 For the first line \eqref{F1_deg} we can use the above result to find 
 \begin{align}
 &\lim_{\lambda \rightarrow 0} \frac{1}{2\ln 2}\sum_{x|\zeta_0(x)\neq 0} \tr{\partial_\lambda^2 \phi_x}\ln \tr{\phi_x} - \mathrm{tr}\left[{\ptr{B}{\partial_\lambda^2 \phi_x}\ln \ptr{B}{\phi_x}}\right]\\& = 
 -\frac{1}{2\ln 2}\sum_{x|\zeta_0(x)\neq 0} \mathrm{tr} \left\{\left(|\zeta_1(x)|^2  - |\zeta_0(x)|^2 \right)\ptr{B}{2 c\phi c^\dag} \ln{\left[ \ptr{B}{\phi}\right]}\right\}.
 \end{align}
The contribution of the next line \eqref{F2_deg} can be rewritten as 
\begin{align}
\frac{1}{2\ln 2}\sum_{x|\zeta_0(x)\neq 0} |\zeta_1(x)|^2 \Bigg[&\bigg|e^{-i\mu_{01}(x)}\tr{c\phi} + e^{+i\mu_{01}(x)}\tr{\phi c^\dag}\bigg|^2\\
    &
    - \sum_{j,k}\frac{\ln \xi_j-\ln\xi_k}{\xi_j-\xi_k}\bigg|e^{-i\mu_{01}(x)}\bra{\xi_j}\ptr{B}{c\phi}\ket{\xi_k} + e^{+i\mu_{01}(x)}\bra{\xi_j}\ptr{B}{\phi c^\dag}\ket{\xi_k}\bigg|^2\Bigg].
\end{align}
The terms in \eqref{F3_deg} are non-trivial. To calculate them, first note that for $\zeta_0(x_0) = 0$, in $\phi_{x_0}$ both the zeroth and first order contribution in $\lambda$ vanish, i.e.  $\phi_{x_0}=|\zeta_1(x_0)|^2 2 c\phi c^\dag \lambda^2 / 2+\mathcal{O}(\lambda^3)$. We consider the case $\zeta_1(x_0)\neq 0$ (the case $\zeta_1(x_0)=0$ gives a vanishing result):
 \begin{align}
\lim_{\lambda \rightarrow 0} & \bigg[\tr{\partial_\lambda^2 \phi_{x_0}}\ln \tr{\phi_{x_0}} - \tr{\ptr{B}{\partial_\lambda^2 \phi_{x_0}}\ln \ptr{B}{\phi_{x_0}}}\bigg]\\
 =&  \lim_{\lambda \rightarrow 0} |\zeta_1(x_0)|^2\bigg(\tr{2 c\phi c^\dag}\ln{\left[\lambda^2|\zeta_1(x_0)|^2\tr{c\phi c^\dag}\right]} - \mathrm{tr} \left\{\ptr{B}{ 2 c\phi c^\dag} \ln{\left[\lambda^2|\zeta_1(x_0)|^2\ptr{B}{ c\phi c^\dag}\right]}\right\}\bigg)\\
  =&  |\zeta_1(x_0)|^2\bigg(\tr{2 c\phi c^\dag}\ln{\left[\tr{c\phi c^\dag}\right]} - \mathrm{tr}\left\{\ptr{B}{ 2 c\phi c^\dag} \ln{\left[\ptr{B}{ c\phi c^\dag}\right]}\right\}\bigg)
 \label{F_first_a0}.
 \end{align}

For the two terms in \eqref{F4_deg} note that $\zeta_0(x_0)=0$ implies $\lim_{\lambda \rightarrow 0} \tr{\phi_{x_0}}=0$ and also $\lim_{\lambda\rightarrow 0}\tr{\partial_\lambda\phi_{x_0}}=0$. From that we find  $\lim_{\lambda\rightarrow 0}\frac{\tr{\partial_\lambda\phi_{x_0}}^2}{\tr{\phi_{x_0}}}=2\tr{\partial_\lambda^2\phi_{x_0}}$. This follows simply by $\phi_x=\phi_x|_{\lambda=0}+\lambda \partial_\lambda \phi_x|_{\lambda=0}+\frac{1}{2}\lambda^2 \partial^2_\lambda \phi_x|_{\lambda=0}+\mc{O}(\lambda^3)$, and $\partial_\lambda \phi_x=\partial_\lambda\phi_x|_{\lambda=0}+\lambda \partial^2_\lambda \phi_x|_{\lambda=0}+\mc{O}(\lambda^2)$, and the L'Hospital's rule. A similar issue arises in the second term of \eqref{F4_deg}, since also $\lim_{\lambda \rightarrow 0}\ptr{B}{\phi_{x_0}}=0$ (as an operator). Then we have 
\begin{align}
\lim_{\lambda\rightarrow 0}\mathrm{tr}\bigg\{{\ptr{B}{\partial_\lambda \phi_{x_0}}R^{-1}_{\ptr{B}{\phi_{x_0}}}\left[\ptr{B}{\partial_\lambda \phi_{x_0}}\right]}\bigg\}=2\mathrm{tr}\bigg\{{\ptr{B}{\partial^2_\lambda \phi_{x_0}}R^{-1}_{\ptr{B}{\partial^2_\lambda \phi_{x_0}}}\left[\ptr{B}{\partial^2_\lambda \phi_{x_0}}\right]}\bigg\}\bigg|_{\lambda=0}=2\tr{\partial^2_\lambda \phi_{x_0}}|_{\lambda=0}.
\end{align}
This can be proven as follows. Consider the expression with the logarithmic derivative: 
  \begin{align}
 \lim_{\lambda\rightarrow 0}\mathrm{tr}\bigg\{{\ptr{B}{\partial_\lambda \phi_{x_0}}R^{-1}_{\ptr{B}{\phi_{x_0}}}\left[\ptr{B}{\partial_\lambda \phi_{x_0}}\right]}\bigg\},
  \end{align}
 where $\phi_{x_0} = \lambda^2 / 2 |\braket{x}{1}|^2 2 c\phi c^\dagger$ and $\partial_\lambda \phi_{x_0} = \lambda |\braket{x}{1}|^2 2 c\phi c^\dagger$ for $|\braket{x}{0}|^2 = 0$ (see definitions above). We can set the matrix $\ptr{B}{2 c\phi c^\dagger} = \sum_i r_i \ket{r_i}\bra{r_i}$ diagonalized in basis $\{\ket{r_i}\}$, such that
\begin{align}
 \ptr{B}{\phi_{x_0}} = \frac{\lambda^2}{2} |\braket{x}{1}|^2 \sum_i r_i \ket{r_i}\bra{r_i},\\ \ptr{B}{\partial_\lambda\phi_{x_0}} = \lambda |\braket{x}{1}|^2 \sum_i r_i \ket{r_i}\bra{r_i}.
\end{align}
Thus the limit simplifies as follows:
\begin{align}
 \lim_{\lambda\rightarrow 0}&\mathrm{tr}\bigg\{{\ptr{B}{\partial_\lambda \phi_{x_0}}R^{-1}_{\ptr{B}{\phi_{x_0}}}\left[\ptr{B}{\partial_\lambda \phi_{x_0}}\right]}\bigg\} =\\=& \lim_{\lambda\rightarrow 0}\mathrm{tr}\Bigg\{{|\braket{x}{1}|^2 \lambda \left(\sum_i r_i \ket{r_i}\bra{r_i}\right) \left[\sum_{j, k}\frac{\ln r_j - \ln r_k}{\lambda^2 /2 (r_j - r_k)} \bra{r_j} \lambda \left(\sum_i r_i \ket{r_i}\bra{r_i}\right) \ket{r_k} \ket{r_j}\bra{r_k}\right]}\Bigg\} = \\
 =& \lim_{\lambda\rightarrow 0} \mathrm{tr}\left[{|\braket{x}{1}|^2 \lambda \left(\sum_i r_i \ket{r_i}\bra{r_i}\right) \frac{2 \lambda}{\lambda^2} \left(\sum_{i} \ket{r_i}\bra{r_i}\right)}\right] = 2 |\braket{x}{1}|^2 \sum_i r_i \equiv 2 \tr{\partial^2_\lambda \phi_{x0}}|_{\lambda=0},
  \end{align}
 where again ${r_i}$ is the set of eigenvalues of matrix $2 c\phi c^\dagger$ and we used the L'Hospital's rule for the expression:
 \begin{align}
     \lim_{r_j\rightarrow r_i} \frac{\ln{r_j} - \ln{r_i} }{r_j - r_i} = \frac{1}{r_i}.
 \end{align}
Thus the terms in \eqref{F4_deg} exactly cancel. In summary, we obtain for a vacuum-resolving case the following expression:
\begin{align}
    \dot{\overline E}_\mathrm{VR}=&  -\frac{1}{2\ln 2}\sum_{x|\zeta_0(x)\neq 0} \left(|\zeta_1(x)|^2  - |\zeta_0(x)|^2 \right) \mathrm{tr} \left\{\ptr{B}{2 c\phi c^\dag} \ln{\left[ \ptr{B}{\phi}\right]}\right\}\\
    &+\frac{1}{2\ln 2}\sum_{x|\zeta_0(x)\neq 0} |\zeta_1(x)|^2 \Bigg[\bigg|e^{-i\mu_{01}(x)}\tr{c\phi} + e^{+i\mu_{01}(x)}\tr{\phi c^\dag}\bigg|^2\\
    &\qquad\qquad\qquad
    - \sum_{j,k}\frac{\ln\xi_j -\ln\xi_k}{\xi_j-\xi_k}\bigg|e^{-i\mu_{01}(x)}\bra{\xi_j}\ptr{B}{c\phi}\ket{\xi_k} + e^{+i\mu_{01}(x)}\bra{\xi_j}\ptr{B}{\phi c^\dag}\ket{\xi_k}\bigg|^2\Bigg]\\
    &+
     \frac{1}{2\ln 2}\sum_{x|\zeta_0(x)= 0}|\zeta_1(x)|^2\bigg(\tr{2 c\phi c^\dag}\ln{\left[\tr{c\phi c^\dag}\right]} - \mathrm{tr}\left\{\ptr{B}{ 2 c\phi c^\dag} \ln{\left[\ptr{B}{ c\phi c^\dag}\right]}\right\}\bigg).
\end{align}
We can further simplify this expression by noting that $\sum_{x|\zeta_0(x)\neq 0}\dots=\sum_{x}\dots - \sum_{x|\zeta_0(x)=0}\dots$. The first line above then simplifies to 
\begin{align}
    +\frac{1}{2\ln 2}\sum_{x|\zeta_0(x)=0}|\zeta_1(x)|^2 \mathrm{tr} \left\{\ptr{B}{2 c\phi c^\dag} \ln{\left[ \ptr{B}{\phi}\right]}\right\}.
    \label{MlogN}
\end{align}
The expression (\ref{MlogN}) seems to be problematic in the case when $\ptr{B}{\phi}$ has zero eigenvalues. However one can notice that, since the jump operator $c$ is local in the B-partition of the system, the matrices $\ptr{B}{\phi}$ and $\ptr{B}{2c\phi c^\dag}$ share the same eigenspace with zero eigenvalue, thus avoiding divergences.

Thus we finally obtain for the vacuum-resolving case:
\begin{align}
   \dot{\overline E}_\mathrm{VR}=& \frac{1}{2\ln 2}\sum_{x|\zeta_0(x)\neq 0} |\zeta_1(x)|^2 \bigg(\bigg|e^{-i\mu_{01}(x)}\tr{c\phi} + e^{+i\mu_{01}(x)}\tr{\phi c^\dag}\bigg|^2\label{F_degen1}\\
    &\qquad\qquad\qquad
    - \sum_{j,k}\frac{\ln\xi_j-\ln\xi_k}{\xi_j-\xi_k}\bigg|e^{-i\mu_{01}(x)}\bra{\xi_j}\ptr{B}{c\phi}\ket{\xi_k} + e^{+i\mu_{01}(x)}\bra{\xi_j}\ptr{B}{\phi c^\dag}\ket{\xi_k}\bigg|^2\bigg)\label{F_degen2}\\
    &+
     \frac{1}{2\ln 2}\sum_{x|\zeta_0(x)= 0}|\zeta_1(x)|^2\Bigg[\tr{2 c\phi c^\dag}\ln{\left[\tr{c\phi c^\dag}\right]} + \mathrm{tr}\left(\ptr{B}{ 2 c\phi c^\dag} \left\{\ln{\left[\ptr{B}{\phi}\right]} - \ln{\left[\ptr{B}{ 2c\phi c^\dag}\right]}\right\}\right)\Bigg].\label{F_degen3}
\end{align}

\subsection{Number measurement}
Let us consider the special case of a number measurement, i.e. the measurement  basis $\{\ket{0}, \ket{1}, \ket{2},\dots\}$. One can immediately see that this corresponds to a vacuum-resolving measurement. We note that for this measurement, when $|\zeta_0(x)|^2 = 0$, it immediately implies that $|\zeta_1(x)|^2 = 1$, and \textit{vice versa}. So that the first terms for $\dot{\overline E}$, Eq.~(\ref{F_degen1}) and Eq.~(\ref{F_degen2}), vanish, as $\zeta_1(x) = 0$, thus the phase-sensitive part of the expression vanishes, as expected for a number measurement setup. The last term, Eq.~(\ref{F_degen3}), stays and gives the expression for a number measurement case:
\begin{align}
    \dot{\overline E}_\mathrm{num}=\tr{c\phi c^\dag}\log_2{\left[\tr{c\phi c^\dag}\right]} + \mathrm{tr}\bigg(\ptr{B}{c\phi c^\dag} \left\{\log_2{\left[\ptr{B}{\phi}\right]} - \log_2{\left[\ptr{B}{ c\phi c^\dag}\right]}\right\}\bigg).
\end{align}
Note that this expression only depends on the state itself and the jump operator. 

\subsection{Homodyne measurement}
Let us now consider a continuous measurement basis in case of an \textit{ideal} homodyne measurement, in which case the measurement basis is the eigenbasis of $b_j(t)e^{i\varphi}+b_j^\dag(t)e^{-i\varphi}$. Since this operator has a continuous spectrum, we replace the summation over $x$ with an integration, $\sum_x \rightarrow \int \,dx$, and the corresponding wave functions of continuous measurement are the wave functions of a harmonic oscillator:
 \begin{align}
    \zeta_0(x) &= \braket{x}{0} = \frac{1}{\pi ^{1/4}} e^{-x^2 /2}, \\
    \zeta_1(x) &= \braket{x}{1} = \frac{\sqrt{2} x e^{-i\varphi}}{\pi^{1/4}} e^{-x^2 /2},\\
    \mu_{01}(x) &= \mu_{01} = \varphi.
 \end{align}
 Clearly, this measurement falls into the  non-vacuum-resolving category (see Subsection \ref{sss:nd}).
 Thus, the expression for the time derivative of entanglement entropy is obtained as [Eqs.~(\ref{F_non-degen1}-\ref{F_non-degen2})]:
 \begin{align}
    \dot{\overline E}_\mathrm{hom}= \frac{1}{2\ln 2}\left(\int \frac{2 x^2}{\sqrt{\pi}} e^{-x^2} \,dx\right) \Bigg[&\bigg|e^{-i\varphi}\tr{c\phi} + e^{+i\varphi}\tr{\phi c^\dag}\bigg|^2 -  \\
    &
    - \sum_{j,k}\frac{\ln\xi_j-\ln\xi_k}{\xi_j-\xi_k}\bigg|e^{-i\varphi}\bra{\xi_j}\ptr{B}{c\phi}\ket{\xi_k} + e^{+i\varphi}\bra{\xi_j}\ptr{B}{\phi c^\dag}\ket{\xi_k}\bigg|^2\Bigg],
\end{align}
where the integral in brackets  is equal to 1, so that we finally obtain:
\begin{align}
    \dot{\overline E}_\mathrm{hom} = \frac{1}{2\ln 2} \Bigg[&\bigg|e^{-i\varphi}\tr{c\phi} + e^{+i\varphi}\tr{\phi c^\dag}\bigg|^2 -  \\
    &
    - \sum_{j,k}\frac{\ln\xi_j-\ln\xi_k}{\xi_j-\xi_k}\bigg|e^{-i\varphi}\bra{\xi_j}\ptr{B}{c\phi}\ket{\xi_k} + e^{+i\varphi}\bra{\xi_j}\ptr{B}{\phi c^\dag}\ket{\xi_k}\bigg|^2\Bigg].
\end{align}
Note that $\dot{\overline E}_\mathrm{hom}$ depends on $\varphi$. In the following we write this dependence explicitly as $\dot{\overline E}_\mathrm{hom}^{\varphi}$.
\subsection{Optimal measurements}
\subsubsection{Non-vacuum-resolving case}
Now let us consider an arbitrary non-vacuum-resolving measurement $\nu$. Note that $\dot{\overline E}_{\nu}\leq 0$, if $c$ has support only in $A$ or $B$, {i.e.} if $c$ is local. Let us define new variables $r$ and $\beta$:
\begin{align}
    \sum_x |\zeta_1(x)|^2e^{-2i\mu_{01}(x)}=r^2e^{-2i\beta}
\end{align}
with a property $\sum_x |\zeta_1(x)|^2=1$. Here $0\leq r\leq 1$, and $0\leq \beta<2\pi$. Note that $r$ and $\beta$ are determined by $\nu$. If $\nu$ is the the ideal homodyne measurement, then we get $r=1$ and $\beta=\varphi$. With these definitions we have
\begin{align}\dot{\overline E}_{\nu}&=\frac{1}{2\ln 2}\bigg[\left(r^2e^{-2i\beta}a_0^2+2|a_0|^2+r^2e^{2i\beta}{a_0^\ast}^2\right)-\sum_i \left(r^2e^{-2i\beta}a_i^2+2|a_i|^2+r^2e^{2i\beta}{a_i^\ast}^2\right)\bigg],
\end{align}
where $a_0 = \tr{c\phi}$ and $a_i = a_{jk} = \bra{\xi_j}\ptr{B}{c\phi}\ket{\xi_k}$, where the index $i$ refers to a double-index $j, k$. Simplifying the latter expression, one can obtain:
\begin{align}
    \dot{\overline E}_{\nu} &= r^2 \dot{\overline E}_{\rm hom}^{(\varphi=\beta)}+\frac{1-r^2}{\ln 2}\left(|a_0|^2-\sum_i|a_i|^2\right),
\end{align}
The expression above is a monotonic function of $r$ in $0\leq r\leq 1$. Thus its minimum is obtained at one of the boundary points, either at $r=0$ or at $r=1$. The options for the minimum value are thus:
\begin{enumerate}[label=(\roman*)]
    \item $\dot{\overline E}_{\nu}=(|a_0|^2-\sum_i|a_i|^2)/{\ln 2}$, or
    \item $\dot{\overline E}_{\nu}=\dot{\overline E}_{\rm hom}^{(\varphi=\beta)}$.
\end{enumerate}

 If $(|a_0|^2-\sum_i|a_i|^2)/{\ln 2}>\dot{\overline E}_{\rm hom}^{(\varphi=\beta)}$, then then homodyne measurement with $\varphi=\beta$ achieves a better $\dot{\overline E}$ then the measurement $\nu$. On the other hand, if $(|a_0|^2-\sum_i|a_i|^2)/{\ln 2}<\dot{\overline E}_{\rm hom}^{(\varphi=\beta)}$, then the homodyne measurement with $\varphi=\beta+\pi/2$ achieves a better $\dot{\overline E}$ then the measurement $\nu$. The last fact can be explained by considering the following relation:
 \begin{align}
     \frac{\dot{\overline E}_{\rm hom}^{(\varphi)}+\dot{\overline E}_{\rm hom}^{(\varphi+\pi/2)}}{2} = \frac{|a_0|^2-\sum_i|a_i|^2}{{\ln 2}}.
 \end{align}
 Since $\dot{\overline E}_{\rm hom}$ is a real value, either $\dot{\overline E}_{\rm hom}^{(\beta)}\leq (|a_0|^2-\sum_i|a_i|^2)/{\ln 2}$ for either $\beta = \varphi$ or $\beta = \varphi + \pi/2$. Thus there exists no non-vacuum-resolving measurement that could give a lower $\dot{\overline E}$ than the ideal homodyne measurement.

\subsubsection{Vacuum-resolving case}
Note that $\dot{\overline E}_{\rm VR}\leq 0$, if $c$ has support only in $A$ or $B$. Let us again define the variables $r$, $\beta$, and $s$:
\begin{align}
    &\sum_{x|\zeta_0(x)\neq 0} |\zeta_1(x)|^2e^{-2i\mu_{01}(x)}=r^2e^{-2i\beta},\\
    &\sum_{x|\zeta_0(x)\neq 0}|\zeta_1(x)|^2=s^2,\\
    &\sum_{x|\zeta_0(x)=0}|\zeta_1(x)|^2=1-s^2.
\end{align}
Here $0\leq r\leq s\leq 1$, and $0\leq \beta<2\pi$. Again, introducing $a_0 = \tr{c\phi}$ and $a_i = a_{jk} = \bra{\xi_j}\ptr{B}{c\phi}\ket{\xi_k}$, we can have:
\begin{align}\dot{\overline E}_{\rm VR}&=\frac{1}{2 \ln 2}\left[\left(r^2e^{-2i\beta}a_0^2+2|a_0|^2+r^2e^{2i\beta}{a_0^\ast}^2\right)-\sum_i \left(r^2e^{-2i\beta}a_i^2+2s^2|a_i|^2+r^2e^{2i\beta}{a_i^\ast}^2\right)\right]+\left(1-s^2\right)\dot{\overline E}_{\rm num},
\end{align}
which can be simplified to:
\begin{align}
    \dot{\overline E}_{\rm VR}&=r^2 \dot{\overline E}_{\rm hom}^{(\varphi=\beta)}+\frac{s^2-r^2}{ \ln 2}\left(|a_0|^2-\sum_i|a_i|^2\right)+\left(1-s^2\right)\dot{\overline E}_{\rm num}.
\end{align}
The above expression is a monotonic function of $r$ in $0\leq r\leq s$ and a monotonic function of $s$ in $0\leq s\leq 1$. Thus the minimum has to be on the boundary of the region $0\leq r\leq s\leq 1$. To proceed we have to distinguish the two cases:
\begin{enumerate}[label=(\roman*)]
    \item $\dot{\overline E}_{\rm hom}^{(\varphi=\beta)}\leq (|a_0|^2-\sum_i|a_i|^2)/\ln 2$, and
    \item $\dot{\overline E}_{\rm hom}^{(\varphi=\beta)}\geq (|a_0|^2-\sum_i|a_i|^2)/\ln 2$.
\end{enumerate}

Let us start with the case (i). In this case $\dot{\overline E}$ is a monotonically decreasing function of $r$. Thus the minimum is achieved at the largest value of $r$, i.e. for $r=s$. Therefore we have:
\begin{align}
    \dot{\overline E}=s^2\dot{\overline E}_{\rm hom}^{(\varphi=\beta)}+(1-s^2) \dot{\overline E}_{\rm num}
\end{align}
This is a monotonic function of $0\leq s\leq 1$. The minimum is thus obtained at the boundary, where the expression is either $\dot{\overline E}_{\rm hom}^{(\varphi=\beta)}$ or $\dot{\overline E}_{\rm num}$.

In the case (ii),  $\dot{\overline E}$ is a monotonically increasing function of $r$. That is, it's minimum is obtained at $r=0$. Therefore we have:
\begin{align}
    \dot{\overline E}=s^2\frac{|a_0|^2-\sum_i|a_i|^2}{\ln 2}+(1-s^2)\dot{\overline E}_{\rm num}
\end{align}
This is again a monotonic function in $s$. It's minimum on $0\leq s\leq 1$ is the smallest of $2(|a_0|^2-\sum_i|a_i|^2)$ and $\dot{\overline E}_{\rm num}$. In case when $\dot{\overline E}_{\rm num}$ is smaller than $2(|a_0|^2-\sum_i|a_i|^2)$, we choose $s = 0$. In the opposite case, note that from the non-vacuum-resolving case above we have:
\begin{align}
    \dot{\overline E}_{\rm hom}^{(\varphi=\beta+\pi/2)}\leq \frac{|a_0|^2-\sum_i|a_i|^2}{\ln 2}.
\end{align}
Thus, no matter what measurement one makes, either the number or homodyne measurement will give the minimal value for $\dot{\overline E}$.

\section{Matrix product state representation and stochastic updates}

\begin{figure}[h]\centering
\includegraphics[width=0.6\linewidth]{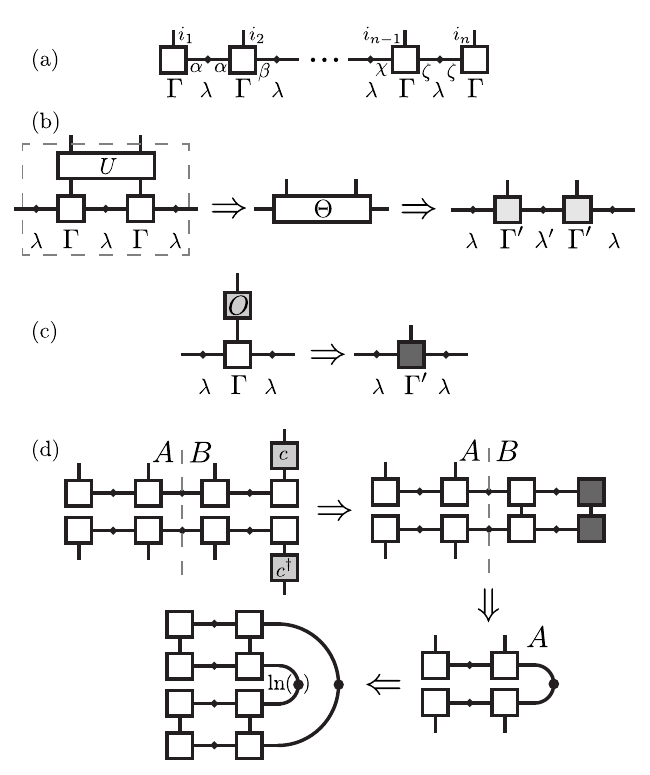}
\caption{Main diagramatic MPS identities used in the paper. (a) $\Gamma\lambda$-form of the state. (b) TEBD-update of two neighbouring sites of the state. (c) Single-site (local) update of the state. (d) Example of MPS contractions needed to calculate the entranglement entropy change rate $\dot{\bar{E}}_j$.
}
\label{fig:SM_Fig2}
\end{figure}

In this section we consider in more detail the evolution of an open many-body quantum system represented by an MPS. We describe the state of the system of $n$ constituents via so-called $\Gamma\lambda$-form of the MPS~(see Ref.~[3] in the main text):
\begin{align}
    &\ket{\phi} = \sum_{i_1, i_2, i_3, ..., i_n} c_{i_1, i_2, i_3, ..., i_n} \ket{i_1, i_2, i_3, ..., i_n},\\
    &\textrm{where } c_{i_1, i_2, i_3, ..., i_n} = \sum_{\alpha, \beta, \gamma, ..., \chi, \zeta} \Gamma_\alpha^{i_1}[1] \lambda_{\alpha\alpha}[1] \Gamma_{\alpha\beta}^{i_2}[2] \lambda_{\beta\beta}[2] \textrm{ }...\textrm{ }\lambda_{\xi\xi}\Gamma_{\xi\zeta}^{i_{n-1}}[n-1] \lambda_{\zeta\zeta}[n-1]\Gamma_{\zeta}^{i_n}[n],
\end{align}
where $i_j$ are the state indices of $j^\mathrm{th}$ constituent, $\Gamma[j]$ and $\lambda[j]$ are a three- and two-dimensional tensors (see Fig.~\ref{fig:SM_Fig2}a). The index $i_j$ in the superscript of the $\Gamma[j]$ tensor is called a physical index and has dimension $d$, indices in the subscript of $\Gamma[j]$ and $\lambda[j]$ tensors correspond to bond indices and have dimension $\chi$.

As discussed in the main text, trajectories are propagated by Trotterizing the coherent and incoherent evolution, leading to alternating layers of state updates as depicted in Fig.~2a of the main text. 


The coherent part is governed by (generally) nearest-neighbour Hamiltonian $H_\mathrm{sys}$. The technique to update the state of the system with such a Hamiltonian is called time-evolving block decimation (TEBD, see Ref.~[4] in the main text). This technique consists in contracting the physical indices of two-local unitary tensors (reshaped unitary matrices) with the corresponding physical indices of the state, as shown in Fig.~2a in the main text. Mathematically one step of this technique is implemented as follows (Fig.~\ref{fig:SM_Fig2}b):
\begin{enumerate}
    \item Apply the unitary tensor to two neighbouring $\Gamma[j]$ and $\Gamma[j+1]$ matrices (Fig.~\ref{fig:SM_Fig2}b, first step):
    \begin{align}
        U\ket{\phi} = &\sum_{i'_{j}, i'_{j+1}}\sum_{i_1, ..., i_j, i_{j+1}, ..., i_n} \sum_{\alpha, ..., \zeta} \Gamma_\alpha^{i_1}[1] \textrm{ }...\textrm{ }U_{i_{j}, i_{j+1}}^{i'_{j}, i'_{j+1}} \lambda_{\theta\theta}[j-1] \Gamma_{\theta\iota}^{i_j}[j] \lambda_{\iota\iota}[j] \Gamma_{\iota\kappa}^{i_{j+1}}[j+1] \lambda_{\kappa\kappa}[j+1] \textrm{ }...\textrm{ }\Gamma_{\zeta}^{i_n}[n] \ket{i_1, i_2, i_3, ..., i_n} = \\
        &\sum_{i_1, ..., i'_j, i'_{j+1}, ..., i_n} \sum_{\alpha, ..., \zeta} \Gamma_\alpha^{i_1}[1] \textrm{ }...\textrm{ }\Theta_{\theta\kappa}^{i'_j, i'_{j+1}}\textrm{ }...\textrm{ }\Gamma_{\zeta}^{i_n}[n] \ket{i_1, i_2, i_3, ..., i_n};
    \end{align}
    \item Perform a singular value decomposition (SVD) on tensor $\Theta$ to obtain updated versions of $\Gamma[j]$, $\lambda[j]$ and $\Gamma[j+1]$ tensors (Fig.~\ref{fig:SM_Fig2}b, second and third steps);
    \item Go to the next pair of matrices $\Gamma[j+2]$ and $\Gamma[j+3]$ and repeat the steps.
\end{enumerate}

This routine is first applied to all the odd pairs of $\Gamma$-matrices and then to all the even pairs.

The incoherent part of the each layer is generated by applying a sequence of single-site operations (denoted by $K_j$ in the main text) to the state of the system. In the MPS framework the application of the operator $K_j$ is achieved by contracting the physical indices with the corresponding index of the $\Gamma[j]$-matrix (Fig.~\ref{fig:SM_Fig2}c, first step), resulting in an update of the  $\Gamma[j]$-matrix (Fig.~\ref{fig:SM_Fig2}c, second step):
\begin{align}
        O\ket{\phi} = &\sum_{i'_{j}}\sum_{i_1, ..., i_j, i_{j+1}, ..., i_n} \sum_{\alpha, ..., \zeta} \Gamma_\alpha^{i_1}[1] \textrm{ }...\textrm{ }O_{i_{j}}^{i'_{j}} \Gamma_{\theta\iota}^{i_j}[j] \textrm{ }...\textrm{ }\Gamma_{\zeta}^{i_n}[n] \ket{i_1, i_2, i_3, ..., i_n} = \\
        &\sum_{i_1, ..., i'_j, ..., i_n} \sum_{\alpha, ..., \zeta} \Gamma_\alpha^{i_1}[1] \textrm{ }...\textrm{ } {\Gamma'}_{\theta\iota}^{i'_j}[j] \textrm{ }...\textrm{ }\Gamma_{\zeta}^{i_n}[n] \ket{i_1, i_2, i_3, ..., i_n}.
\end{align}
    
This procedure is applied to all sites, after which the MPS is not in canonical form and not normalized. At the end of the stochastic layer, the MPS is renormalized and the canonical $\Gamma\lambda$-form is reinstated by TEBD technique with an identity unitary matrix (Fig.~\ref{fig:SM_Fig2}c, last step, see Ref.~[3] in the main text).

The choice of single-site operations $K_j$ depends on a specific type of unravelling. In the EOQT method that we propose in the main text, the choices of $K_j$ are time-, trajectory- and site-dependent, since they are based on the minimization of the EAEE change rate $\dot{\bar{E}}_j$ [for example, see Eqs.~(10) and~(11) in the main text]. Note that $\dot{\bar{E}}_j$ can be efficiently calculated within the framework of MPSs. To see this, consider, for example, the following term from Eq.~(10) in the main text: 
\begin{align}
    \mathrm{tr}\bigg\{\ptr{B}{c\phi c^\dagger} \ln\left[\ptr{B}{c\phi c^\dagger}\right]\bigg\},
    \label{eq:example}
\end{align}
the diagram of which is shown in Fig.~\ref{fig:SM_Fig2}d. First, the MPS representation of a density matrix $\phi$ is constructed, to which two operators are applied to obtain $c\phi c^\dagger$ (Fig.~\ref{fig:SM_Fig2}d, step one). Then the partial trace over $B$ is taken to obtain $\ptr{B}{c\phi c^\dagger}$ (Fig.~\ref{fig:SM_Fig2}d, step two and three). Finally, the logarithm of eigenvalues of $\ptr{B}{c\phi c^\dagger}$ is taken and tracing over $A$ of two objects, $\ptr{B}{c\phi c^\dagger}$ and $\ln \left[\ptr{B}{c\phi c^\dagger}\right]$, is performed to obtain Eq.~\eqref{eq:example}. All the other terms entering $\dot{\bar{E}}_j$ can be calculated in a similar way.

\section{Numerical case studies} 
In this section we apply the EOQT method as well as other QT methods to several examples in order to illustrate the convergence properties on concrete physical models. 
\subsection{Convergence with number of trajectories}

\begin{figure}[h]\centering
\includegraphics[width=\linewidth]{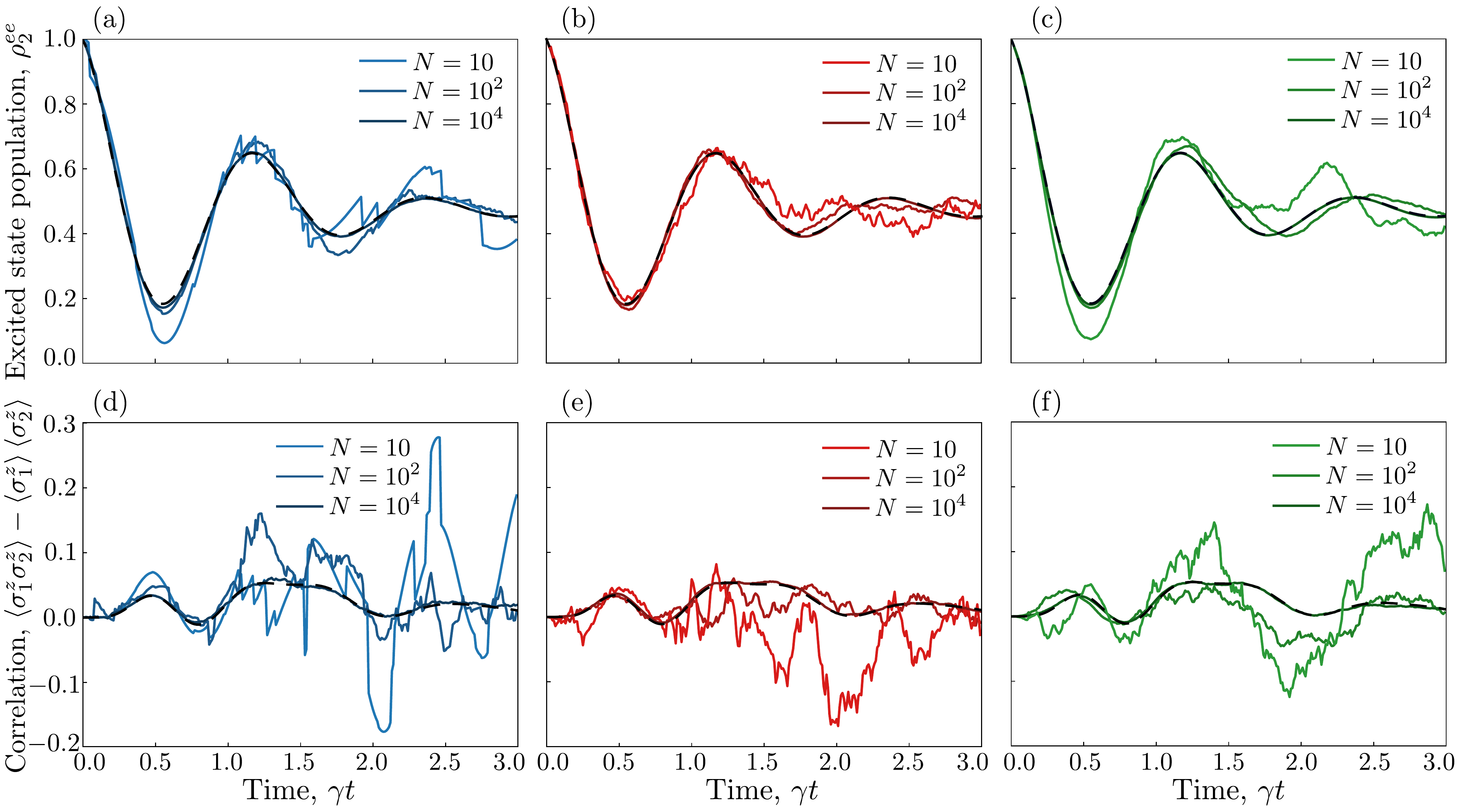}
\caption{Convergence of various QT approaches to the ME solution on the example of the nearest-neighbour Ising model with dephasing [see Eq.~\eqref{eq:H_Ising}]. We show the evolution of the excited state population of the second atom in the chain, $\rho^{ee}_2(t)$, and correlation $\left<\sigma^z_1 \sigma^z_2\right> - \left<\sigma^z_1\right>\left<\sigma^z_2\right>$, in case of the quantum jump method [(a) and (d)], the quantum state diffusion method [(b) and (e)] and the EOQT method [(c) and (f)], for $N=10$, $N=10^2$ and $N=10^4$ trajectories in each case. The black dashed line indicates the exact ME solution, which coincides with the ensemble averages for $N=10^4$ trajectories. The initial state is the fully polarized state with all atoms in the state $\ket{1}$, and the system parameters are $g/\gamma = 2.5$, $J/\gamma = 0.5$, $h/\gamma = -0.5$, and $n = m = 4$.
}
\label{fig:SM_Fig3}
\end{figure}
\begin{figure}[h!]\centering
\includegraphics[width=0.32\linewidth]{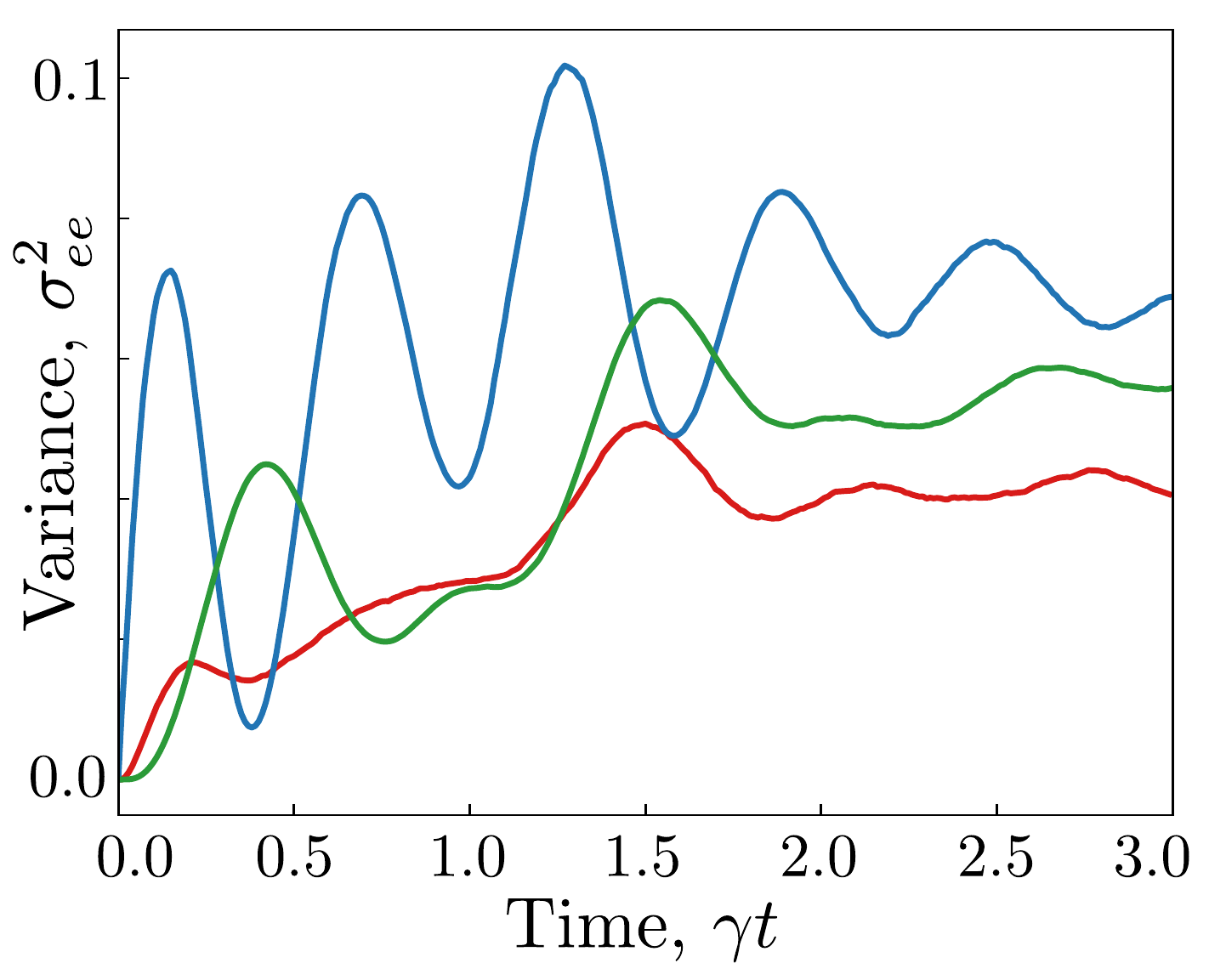}
\caption{Comparison of noise properties of existing QT methods. Stochastic fluctuations of mean excited state population of the second atom in the chain, $\sigma^2_{ee}(t)$, in case of quantum jump method (blue), quantum state diffusion method (red) and EOQT method (green) calculated by sampling $N=10^4$ trajectories for each method. One can notice that EOQT method inherits noise properties from both number and homodyne measurement. The system with a number of constituents $n=4$ is initialized in a product state $\ket{\psi} = \ket{1111}$. Parameters of the system are: $g/\gamma = 2.5$, $J/\gamma = 0.5$, $h/\gamma = -0.5$.
}
\label{fig:SM_Fig4}
\end{figure}
We first apply various QT methods to the paradigmatic example of the open  1D Ising models. This serves, first, to illustrate numerically that the all QT methods indeed converge to the exact solution of the ME and, second, to provide an analysis of the statistical convergence properties on a well-studied example. Specifically, we consider an open 1D spin chain described by the ME $\dot\rho=-i[H_{\rm sys},\rho]+\gamma\sum_j(c_j\rho c_j^\dag-\frac{1}{2}c_j^\dag c_j \rho-\frac{1}{2}\rho c_j^\dag c_j)$, whose coherent part is given by a nearest-neighbor Ising interaction with transverse and longitudinal fields:
\begin{align}
    H_\mathrm{sys} =  \sum_{j=1}^{n} \left[h \sigma_j^z - g \sigma_j^x\right] + \sum_{j=1}^{n-1} J\sigma_j^z\sigma_{j+1}^z, 
    \label{eq:H_Ising}
\end{align}
and the incoherent part is generated by the operators $c_j=\ket{0}_j\,\bra{1}$.

In Fig.~\ref{fig:SM_Fig3} we show exemplary results obtained from the exact integration of the ME and compare it with the QT methods, where the trajectories are produced by continuous number measurements [Fig.~\ref{fig:SM_Fig3}a and~\ref{fig:SM_Fig3}d], quantum state diffusion method that corresponds to homodyne measurements [Fig.~\ref{fig:SM_Fig3}b and~\ref{fig:SM_Fig3}e], and the EOQT method that utilizes the optimal propagator explained in the main text [Fig.~\ref{fig:SM_Fig3}c and~\ref{fig:SM_Fig3}f]. As one can see, all of the possible trajectories converge to the ME solution (black dashed lines in Fig.~\ref{fig:SM_Fig3}), in case when the number of trajectories $N$ is sufficiently large.
 For all trajectory methods, the statistical error of a linear observable $O$ decreases with the number of trajectories as $\sim C_O/\sqrt{N}$. The prefactor $C_O$ depends in general on the observable of interest, but, more importantly for the present discussion, also on the QT method. This can be quantified by the variance of expectation values of interest across trajectories, $C_O^2\sim \sigma_{O}^2=\sum_i p^{(i)}\bra{\phi^{(i)}}O\ket{\phi^{(i)}}^2-\lr{\sum_i p^{(i)} \bra{\phi^{(i)}}O\ket{\phi^{(i)}}}^2$, shown in Fig.~\ref{fig:SM_Fig4}. This plot illustrates that, even though all types of the trajectories converge to the ME solution, the convergence rates differ for various environment monitoring methods.  

\subsection{Convergence with bond dimension}

\begin{figure}[h]\centering
\includegraphics[width=0.75\linewidth]{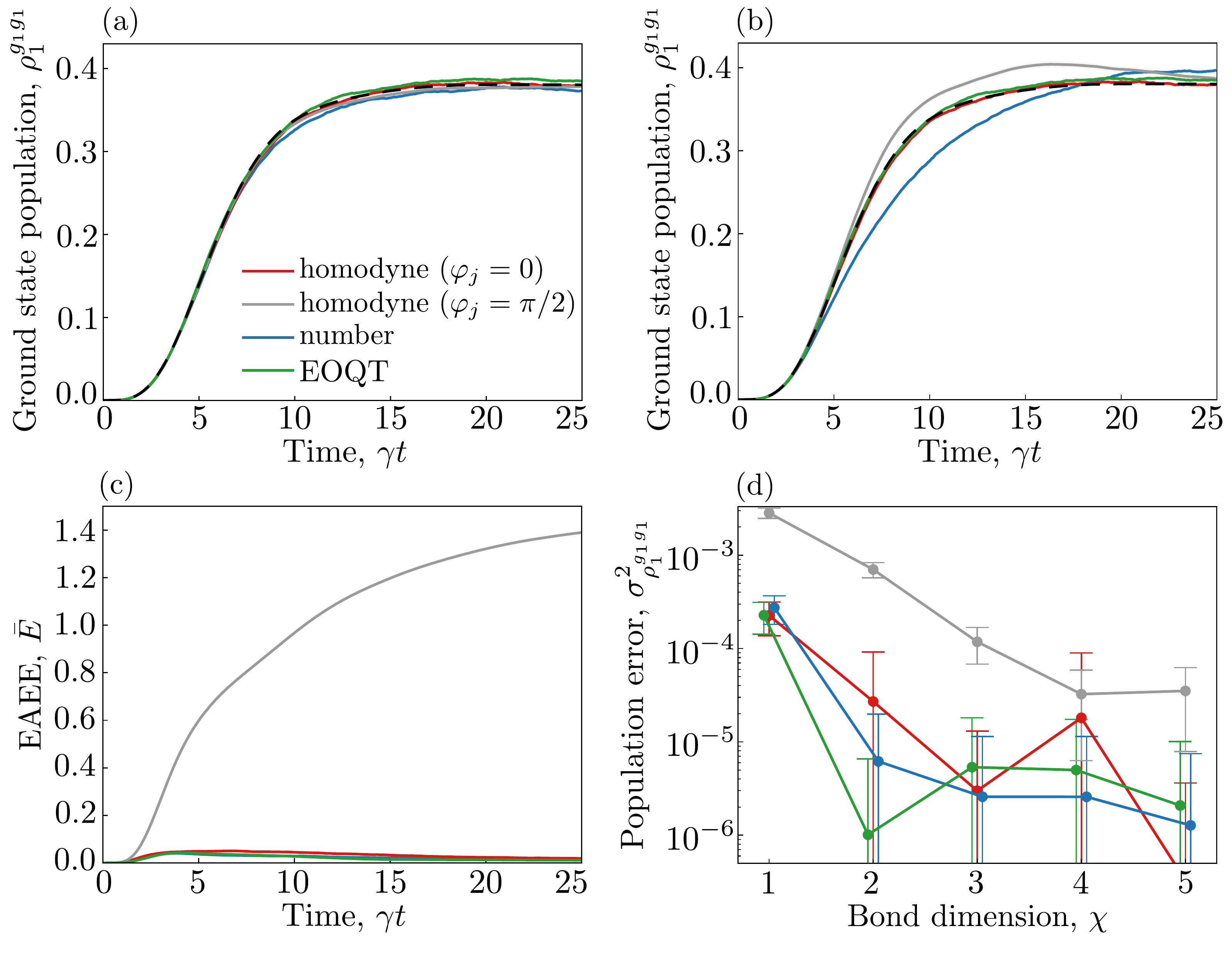}
\caption{Comparison between different unravellings in terms of their ability to solve the ME at a given bond dimension $\chi$. Population of $\ket{g_1}$ state of the first atom in the chain, produced by homodyne measurements with phases $\varphi_j = 0$ (red solid lines) and $\varphi_j = \pi/2$ (grey solid lines), number measurements (blue solid lines) and EOQT-measurements (green solid lines) for bond dimension (a) $\chi = 9$ and (b) $\chi = 1$. The dashed line corresponds to the ME solution. One can see that trajectories with $\varphi_j = \pi/2$ fail to solve the ME, while trajectories with $\varphi_j = 0$ as well as optimal trajectories do not. (c) EAEE of the ensembles produced by homodyne measurements with phases $\varphi_j = 0$ (red solid line) and $\varphi_j = \pi/2$ (grey solid line), number measurements (blue solid line) and EOQT-measurements (green solid line) for $\chi = 9$. (d) The error in the population of $\ket{g_1}$ state of the first atom in the chain, $\sigma^2_{\rho_1^{g_1 g_1}}$, obtained for the time instant $\gamma t = 17.5$, produced by homodyne measurements with phases $\varphi_j = 0$ (red solid line) and $\varphi_j = \pi/2$ (grey solid line), number measurements (blue solid line) and EOQT-measurements (green solid line). In this figure: Number of atoms and dephasing channels $n = m = 4$~[for (a), (b) and (c)] and $n = m = 20$~[for (d)], number of trajectories $N = 10^4$, $\Omega_1 = \Omega_2 = 0.5\gamma$ and $V = \gamma$, where $\gamma$ is the decoherence rate.
}
\label{fig:SM_Fig5}
\end{figure}

The main idea underlying the design of our EOQT method is the expectation that it allows to solve the ME with trajectories that require a reduced bond dimension (compared to other QT methods). To illustrate this feature on a concrete physical model, we choose to consider in the following a model of an open 1D chain of 3-level atoms (with states denoted $\ket{g_1}, \ket{g_2} $ and $\ket{r}$). We already demonstrated this in the main text on the example of an open RBC (Fig.~3 of the main text), where some QT methods lead to volume-law-entangled trajectories (MPSs with finite bond dimension will fail to represent these trajectories accurately), while other methods, including EOQT, generate area-law-entangled trajectories, allowing for an accurate representation by an MPS with finite bond dimension. While the open RBC allows to elegantly demonstrate this feature semi-analytically, other aspects regarding the convergence with bond dimension are less transparent in this model: Since the open RBC quickly generates a maximally mixed state, uncontrolled errors in the representation of individual trajectories due to bond dimension limitations often do not result in errors in physically relevant observables, since the state is maximally mixed. We therefore choose to illustrate the difference of various QT schemes regarding the convergence of physically relevant observables with the bond dimension on a model that does not suffer from this shortcoming. The coherent part of the evolution is given by
\begin{align}
    H_\mathrm{EIT} = - \sum_{j=1}^{n} \left[\frac{\Omega_1}{2}\left(\sigma^{+}_{1,j}+\sigma^-_{1,j}\right) + \frac{\Omega_2}{2}\left(\sigma^+_{2,j}+\sigma^-_{2,j}\right) \right] + \sum_{j=1}^{N-1} V \sigma^z_{1,j}\otimes \sigma^z_{1,j+1},
\end{align}
where $\sigma^-_{i,j}=\ket{g_i}_j\bra{r}$ is the spin-lowering operator of the transition $\ket{g_i} \leftrightarrow \ket{r}$ of atom $j$, and $\sigma^+_{i,j}=(\sigma^-_{i,j})^\dag$. The incoherent part is induced by jump operators of the form $c_j=\ket{r}_j\bra{r}$ (i.e. dephasing channels). This is a physically relevant model describing a Rydberg atom array driven under conditions of electromagnetically induced transparency (EIT). In Fig.~\ref{fig:SM_Fig5} we show the dynamics of this chain obtained from an exact solution using the ME and compare it with various QT schemes using MPSs with different values of the maximum bond dimension $\chi$. As one can see, all QT schemes reproduce the result of the ME in the limit of large bond dimension (as expected). However, when the maximum bond dimension is reduced, some QT methods result in significant errors, while others, including the EOQT method, still produce accurate results.

\section{Further details on open Random Brownian circuits}

In this section we provide further details on some points in the analysis of the RBC. To start, consider the stochastic propagator associated with the homodyne measurement for a jump operator $c_j = \sigma^z_j$ and equal decay rates $\gamma_j = \gamma$ [Eq.~(5) in the main text]:
\begin{align}
    K_j=&e^{-\gamma dt/2} +\sqrt{\gamma} \sigma^z_j e^{i\varphi_j}\left[2 \sqrt{\gamma} \left<\sigma^z_j\right>(t) \cos{\varphi_j} dt + dW_j(t) \right],\label{eq:propag_hom}
\end{align}
where $\left<\sigma^z_j\right>(t) = \bra{\phi_j^{(k)}(t)} \sigma^z_j \ket{\phi_j^{(k)}(t)}$ ($k$ is the trajectory index) and $d W_j (t)$ is a Wiener increment. In the discussion of the RBC in the main text we use a different form of the propagator Eq.~\eqref{eq:propag_hom}, given by $\exp\left\{e^{i\varphi_j} \sqrt{\gamma}\sigma^z_j \left[2 \sqrt{\gamma} \left<\sigma^z_j\right>(t) \cos{\varphi_j} dt + dW_j(t)\right]\right\}$. To show the equivalence between the two propagators, let us consider the expansion:
\begin{align}
    &\exp\left\{e^{i\varphi_j} \sqrt{\gamma}\sigma^z_j \left[2 \sqrt{\gamma} \left<\sigma^z_j\right>(t) \cos{\varphi_j} dt + dW_j(t)\right]\right\}\\
    &= \mathbb{1} + e^{i\varphi_j} \sqrt{\gamma}\sigma^z_j \left[2 \sqrt{\gamma} \left<\sigma^z_j\right>(t) \cos{\varphi_j} dt + dW_j(t)\right] + \frac{1}{2} e^{i 2\varphi_j} dW^2_j(t) \gamma \mathbb{1} + \mathcal{O}\left(dt^{3/2}\right)\\
    &= \mathbb{1} + e^{i\varphi_j} \sqrt{\gamma}\sigma^z_j \left[2 \sqrt{\gamma} \left<\sigma^z_j\right>(t) \cos{\varphi_j} dt + dW_j(t)\right] - \frac{\gamma dt}{2} \mathbb{1} + \frac{\gamma}{2} \left[dt + e^{i 2\varphi_j} dW^2_j(t)\right]\mathbb{1} + \mathcal{O}\left(dt^{3/2}\right)
    \\&= 
    K_j + \frac{\gamma}{2} \left[dt + e^{i 2\varphi_j} dW^2_j(t)\right] K_j+ \mathcal{O}\left(dt^{3/2}\right) \\
    &= K_j \exp{\left\{\frac{\gamma}{2} \left[dt + e^{i 2\varphi_j} dW^2_j(t)\right]\right\}} + \mathcal{O}\left(dt^{3/2}\right).
\end{align}
The two propagators are thus equivalent up to some higher-order terms vanishing in the $dt\rightarrow 0$ limit and a factor $f = \exp{\left[\gamma dt/2 + \gamma e^{i 2\varphi_j} dW^2_j(t) / 2\right]}$. The factor $f$ is a complex number that only affects the normalization and the global phase of the state. Since after each update the state in normalized, this factor is irrelevant. Therefore the propagation with $K_j$ is equivalent to propagation with $\exp\left\{e^{i\varphi_j} \sqrt{\gamma}\sigma^z_j \left[2 \sqrt{\gamma} \left<\sigma^z_j\right>(t) \cos{\varphi_j} dt + dW_j(t)\right]\right\}$.

\end{document}